\documentclass[journal,transmag]{IEEEtran}

\usepackage{graphicx}%
\usepackage{multirow}%
\usepackage{amsmath,amssymb,amsfonts}%
\usepackage{amsthm}%
\usepackage{mathrsfs}%
\usepackage{xcolor}%
\usepackage{textcomp}%
\usepackage{manyfoot}%
\usepackage{booktabs}%
\usepackage{algorithm}%
\usepackage{algorithmicx}%
\usepackage{algpseudocode}%
\usepackage{listings}%
\usepackage{enumitem}
\usepackage{float}
\usepackage{ragged2e}
\usepackage{parskip}
\usepackage{amsmath}
\usepackage{amssymb}
\usepackage{dblfloatfix}
\newcommand{\mb}{\mathbf{m}}

\def\eb   {{\bf{e}}}

\def\Heff {{\bf{H}_{\rm eff}}}
\def\Hbias {{\bf{H}_{\rm bias}}}
\def\Hdemag {{\bf{H}_{\rm demag}}}
\def\Hani {{\bf{H}_{\rm ani}}}
\def\Hexch {{\bf{H}_{\rm exch}}}
\def\HDMI {{\bf{H}_{\rm DMI}}}

\def\mb   {{\bf{m}}}
\def\Mb   {{\bf{M}}}
\def\Nb   {{\bf{N}}}
\def\nb   {{\bf{n}}}
\def\rb   {{\bf{r}}}

\def\MRb {\boldsymbol{\mathcal{R}}}

\newcommand*\wraptt[1]{{\small\texttt{\expandafter\ttbreak\detokenize{#1}\relax}}}
\newcommand*\ttbreak[1]{\ifx\relax#1\else
  \expandafter\ifx\string_#1\string_\allowbreak\else#1\fi
  \expandafter\ttbreak\fi}

\begin{document}

\title{MagneX: A High-Performance, GPU-Enabled, Data-Driven Micromagnetics Solver for Spintronics}

\author{%
\IEEEauthorblockN{%
Andy Nonaka\IEEEauthorrefmark{1},
Yingheng Tang\IEEEauthorrefmark{1},
Julian C.\ LePelch\IEEEauthorrefmark{1},
Prabhat Kumar\IEEEauthorrefmark{1},
Weiqun Zhang\IEEEauthorrefmark{1}, \\
Jorge A.\ Mu\~noz\IEEEauthorrefmark{1,2},
Christian Fernandez\mbox{-}Soria\IEEEauthorrefmark{1,2},
C\'esar D\'iaz\IEEEauthorrefmark{1,2},
David J.\ Gardner\IEEEauthorrefmark{3}, and
Zhi Jackie Yao\IEEEauthorrefmark{1}}%
\IEEEauthorblockA{\IEEEauthorrefmark{1}Applied Mathematics and Computational Research Division, Lawrence Berkeley National Laboratory, Berkeley, CA, USA}%
\IEEEauthorblockA{\IEEEauthorrefmark{2}The University of Texas at El Paso, El Paso, TX, USA}%
\IEEEauthorblockA{\IEEEauthorrefmark{3}Center for Applied Scientific Computing, Lawrence Livermore National Laboratory, Livermore, CA, USA}%
\thanks{Corresponding author: Z.\ Yao (email: \texttt{jackie\_zhiyao@lbl.gov}).}%
}


\IEEEtitleabstractindextext{%
\begin{abstract}
In order to comprehensively investigate the multiphysics coupling in spintronic devices, it is essential to parallelize and utilize GPU-acceleration to address the spatial and temporal disparities inherent in the relevant physics.
Additionally, the use of cutting-edge time integration libraries as well as machine learning (ML) approaches to replace and potentially accelerate expensive computational routines are attractive capabilities to enhance modeling capabilities moving forward.
Leveraging the Exascale Computing Project software framework AMReX, as well as SUNDIALS time-integration libraries and python-based ML workflows, we have developed an open-source micromagnetics modeling tool called MagneX.
This tool incorporates various crucial magnetic coupling mechanisms, including Zeeman coupling, demagnetization coupling, crystalline anisotropy interaction, exchange coupling, and Dzyaloshinskii-Moriya interaction (DMI) coupling.
We demonstrate the GPU performance and scalability of the code and rigorously validate MagneX's functionality using the mumag standard problems and widely-accepted DMI benchmarks.
Furthermore, we demonstrate the data-driven capability of MagneX by replacing the computationally-expensive demagnetization physics with neural network libraries trained from our simulation data.
With the capacity to explore complete physical interactions, this innovative approach offers a promising pathway to better understand and develop fully integrated spintronic and electronic systems.
\end{abstract}

}

\maketitle

\IEEEdisplaynontitleabstractindextext

\IEEEpeerreviewmaketitle

\section{Introduction}\label{sec:intro}

\IEEEPARstart{T}{he} dynamic behavior of magnetization in ferromagnetic materials underpins a wide range of modern technologies, including spintronic logic~\cite{xie2025emerging, ahn20202d}, non-volatile magnetic memory~\cite{sousa2005non}, and high-sensitivity magnetic sensors~\cite{yuan2021extremely}. A fundamental understanding of these phenomena is essential for the continued advancement of nanoscale information processing and storage. At the core of micromagnetic modeling lies the Landau-Lifshitz-Gilbert (LLG) equation~\cite{lakshmanan2011fascinating}, a nonlinear partial differential equation that governs the time evolution of the magnetization vector field in response to effective magnetic fields. Accurate and efficient numerical solutions to the LLG equation are indispensable for capturing the complex dynamics of magnetic microstructures, including the formation and transformation of magnetic domains~\cite{song2021direct, matsubara2004computer}, the motion and pinning of domain walls~\cite{tan2021domain, kaappa2024magnetic}, and the stability of topologically nontrivial configurations such as skyrmions~\cite{de2023skyrmion} and magnetic vortices~\cite{tazes2024efficient}. These micromagnetic simulations enable both predictive modeling of device behavior and rational selection of novel materials with tailored magnetic responses.
Due to the multiscale nature of micromagnetics -- with spatial features spanning from nanometers to sub-millimeter scales and characteristic time scales spanning femtoseconds to microseconds -- LLG solvers must contend with high computational demands. In response, several high-performance micromagnetics solvers have been developed, such as OOMMF~\cite{donahue1999oommf}, MuMax and MuMax3~\cite{vansteenkiste2011mumax, vansteenkiste2014design}, with the latter leveraging GPU acceleration to improve scalability and runtime efficiency. More recent efforts include MagTense~\cite{bjork2021magtense}, FastMag~\cite{chang2011fastmag} and GPU-based solvers capable of multi-GPU execution~\cite{lepadatu2023accelerating}, as well as explorations of parallel-in-time integration methods to further extend parallelism~\cite{kraft2024parallel}.

While existing solvers have significantly advanced the state of micromagnetic simulation, they are constrained by key limitations that hinder their applicability to large-scale, multiscale, and multiphysics problems. These solvers typically rely on explicit, small-time-step integration methods, which become inefficient for stiff physical processes. Moreover, they lack built-in support for machine learning (ML) surrogate integration, limiting their extensibility toward hybrid modeling or differentiable simulation workflows. A critical scalability bottleneck in existing solvers is their limited ability to exploit multi-GPU architectures at scale. For example, MuMax3 is optimized for single-GPU execution with only limited or experimental support for distributed parallelism, while OOMMF remains CPU-based by default, with GPU acceleration available only through separate extensions or add‑on packages. As a result, neither tool is well suited for large-scale, multi-GPU simulations on modern high-performance computing systems. 


To overcome these challenges, we develop MagneX, a modular, GPU-enabled micromagnetics framework that combines multirate time integration, machine learning module integration, and scalable multi-GPU execution, all built atop the AMReX software ecosystem~\cite{AMReX_JOSS} optimized for exascale computing.
MagneX is open-source and designed to address the dual challenges of fidelity and performance in large-scale magnetic simulations. 
It combines GPU-accelerated computation with advanced time integration methods provided by the SUNDIALS suite~\cite{hindmarsh2005sundials, gardner2022sundials} to support accurate and stable long-time simulations. Furthermore, we incorporate ML surrogates to reduce the cost of the more computationally-expensive physical process, i.e. the calculation of the demagnetization field. This hybrid strategy enables fast, scalable, and high-fidelity modeling of complex magnetization dynamics, supporting both fundamental studies and the design of next-generation spintronic devices.

In the sections that follow, we first describe the architecture and mathematical formulation of MagneX, including its mathematical derivations, support for multirate time integration via the SUNDIALS suite, and its use of AMReX for multi-GPU parallel mesh management. 
We then verify the numerical correctness of the solver through various micromagnetic benchmarks, including the $\mu$MAG Standard Problems and simulations of micromagnetic vortex and skyrmion dynamics with Dzyaloshinskii-Moriya interaction (DMI). 
After establishing baseline solver accuracy, we integrate a neural network (NN) surrogate to replace the demagnetization field computation and evaluate its effectiveness within the simulation framework. We conclude with a discussion of scalability, current limitations, and future directions for hybrid micromagnetic modeling.

\section{Overview of MagneX}\label{sec:overview}
\begin{figure}[!t]
    \centering
    \includegraphics[width=0.5\textwidth]{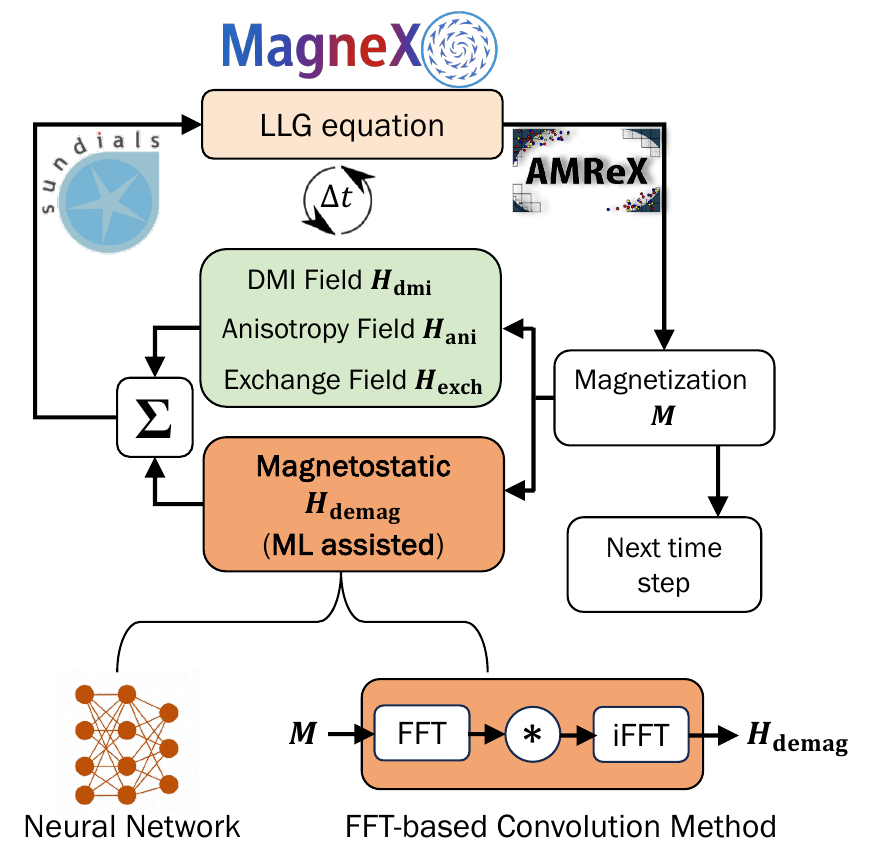}
    \caption{\textbf{Architecture of the MagneX micromagnetic solver.} The Landau–Lifshitz–Gilbert (LLG) equation is evolved using a multirate time integration scheme (SUNDIALS), which treats fast (exchange, anisotropy, DMI) and slow (demagnetization) components on separate time scales. Spatial discretization and parallelization are handled via the AMReX framework. The demagnetizing field $\Hdemag$ can be computed either via FFT-based spectral convolution or approximated by a neural network surrogate. This hybrid design enables efficient, scalable, and extensible simulations on modern multi-GPU architectures.}
    \label{fig:fig1_magnex_architecture}
\end{figure}

Figure~\ref{fig:fig1_magnex_architecture} provides a schematic overview of the MagneX solver architecture. The framework combines spatial decomposition via AMReX, multirate time integration using SUNDIALS, and modular computation of effective magnetic field components. The demagnetization term can be computed either via Fast Fourier Transform (FFT) based convolution method or replaced by a neural network surrogate. We now formalize the underlying model by presenting the governing LLG equation and its constituent field terms.

\subsection{Mathematical Model and Effective Field Components}

The core mathematic model of the MagneX simulation framework is based on the LLG equation~\cite{landau1992theory}, which describes the time evolution of the magnetization vector field $\Mb(\rb, t)$ under the influence of an effective magnetic field $\Heff$:
\begin{equation}
\frac{\partial\Mb}{\partial t} = \mu_0 \gamma_\mathrm{L} (\Mb \times \Heff) + \frac{\alpha \mu_0 \gamma_\mathrm{L}} {M_s} \Mb \times (\Mb \times \Heff) \equiv \MRb.
\label{eq:LLG}
\end{equation}

Here, $\Mb = (M_x, M_y, M_z)$ is the magnetization vector, $M_s$ a space-dependent saturation magnetization, $\mu_0$ the vacuum permeability, and $\alpha$ the Gilbert damping parameter, where $\gamma_L = \gamma / (1 + \alpha ^2)$. The gyromagnetic ratio $\gamma$ equals $-1.759\times 10^{11}$~C/kg for electrons (the minus sign indicates opposite alignment between magnetic moment and electron spin). $\alpha$ typically falls between $10^{-1}$ and $10^{-4}$, with its positive value corresponding to decaying precession of $\Mb$. This parameter encompasses various damping mechanisms such as lattice vibrations, spin-spin relaxation, and spin-orbit coupling. We adopt the LLG equation to describe magnetization evolution since it adequately represents damping phenomena and allows simple mathematical inclusion in the magnetic susceptibility formulation.

The two terms in equation (1) can be readily distinguished. The first, proportional to $\Mb \times \Heff$, constitutes a torque that produces an angular momentum vector tangent to the precessional orbit of $\Mb$. The second, proportional to $\Mb \times (\Mb \times \Heff)$, provides a damping vector oriented perpendicularly to $\Mb$ and directed inward toward the center of its precession path. Note that analytically, the magnitude of magnetization does not change, because we assume the magnetization is saturated -- a valid approximation for both radio-frequency and memory applications.

The effective magnetic field, $\Heff$, drives the magnetic spins and can be decomposed into several terms accounting for the contributions from external bias $\Hbias$, demagnetization $\Hdemag$, crystal anisotropy $\Hani$, exchange coupling $\Hexch$, and Dzyaloshinskii-Moriya interaction (DMI) coupling $\HDMI$:
\begin{equation}
\Heff = \Hbias + \Hdemag + \Hani + \Hexch + \HDMI.
\label{eq:Heff}
\end{equation}

Each term in equation (\ref{eq:Heff}) corresponds to a physically meaningful interaction, implemented as a modular component in the MagneX framework:

\begin{itemize}
\item \textbf{External bias field} $\Hbias$ is a user-defined time-variant or -invariant applied field.

\item \textbf{Demagnetization field} $\Hdemag$ is the self-generated magnetic field resulting from magnetostatic (dipolar) interactions within the magnetized material, calculated via convolution of the magnetization $\Mb$ with the demagnetization tensor $\Nb$.
\begin{equation}
    \Hdemag(\rb) = \int_{\Omega}\Nb(\rb-\rb')\Mb(\rb')d\rb',
\end{equation}
and associated energy density is
\begin{equation}
    e_{\rm demag} = -\frac{\mu_0}{2}(\mb \cdot \Hdemag),
    \label{energy_demag}
\end{equation}
where $\mb=\Mb/M_s = (m_x,m_y,m_z)$ is the normalized magnetization vector.

\item \textbf{Anisotropy field} $\Hani$ describes the effects due to the local orientation of the crystal lattice, and is equal to
\begin{equation}
\Hani = \frac{2K_u}{\mu_0 M_s^2}(\Mb\cdot\eb_K)\eb_K,
\end{equation}
where $K_u$ is the anisotropy constant and $\eb_K$ is a unit vector along the easy axis of the crystal. In a system with uniaxial anisotropy, we can determine the energy density as a product of $K_u$ with the average of the angles between the magnetic moments and the easy axis \cite{d2004nonlinear}:

\begin{equation}
e_{\rm ani} = K_u[1 - (\eb_K \cdot \mb)^2].
\label{energy_ani}
\end{equation}

\item \textbf{Exchange field} $\Hexch$ models the tendency for the magnetization to align with neighboring magnetization, which is why it is also called Heisenberg or symmetric exchange coupling. It is given by
\begin{equation}
\Hexch = \frac{2}{\mu_0 M_s^2}\nabla\cdot A\nabla\Mb,
\end{equation}
with associated energy density
\begin{equation}
 e_{\rm exch} = A[(\nabla m_x)^2 + (\nabla m_y)^2 + (\nabla m_z)^2]
 \label{energy_exch}
\end{equation}
where $A$ is the exchange constant. 

\item \textbf{DMI field} $\HDMI$  originates from spin-orbit coupling and stabilizes chiral magnetic structures. Unlike the symmetric Heisenberg exchange that favors parallel alignment of neighboring moments, DMI is an antisymmetric exchange that favors perpendicular alignment, corresponding mathematically to a cross-product term. 
The interfacial form for thin films reads~\cite{rohart2013skyrmion}:
\begin{equation}
\HDMI = -\frac{2D}{\mu_0 M_s^2} \left[ (\nabla \cdot \Mb)\, \eb_z - \nabla M_z \right],
\end{equation}
where $D < 0$ is the DMI constant. For numerical implementations, a common vector identity form is:
\begin{equation}
\frac{\partial \Mb}{\partial x_k} = -\frac{D}{2A} (\eb_z \times \eb_k) \times \Mb,
\end{equation}
which expresses the DMI contribution in terms of magnetization gradients along spatial directions.
\end{itemize}

\subsection{Spatial Discretization and Demagnetization Solver}

MagneX employs a uniform spatial discretization using cubic grid cells. The computational domain is mapped onto a structured mesh managed by the AMReX framework, with each physical quantity defined at cell centers. Spatial derivatives in the exchange and DMI terms are discretized using standard second-order central difference stencils.

Boundary conditions depend on the physics enabled. In the absence of DMI, zero-flux Neumann conditions are imposed such that $\partial\Mb/\partial\nb = 0$ at material interfaces. When DMI is active, an inhomogeneous Neumann boundary condition is applied directly, consistent with established treatments of chiral boundary effects in micromagnetic systems.

Demagnetization fields are computed via an FFT-based convolution. We follow the method of Nakatani et al.~\cite{nakatani1989direct} and Hayashi et al.~\cite{hayashi1996calculation}, which uses a real-space demagnetization tensor $\Nb$ defined on an extended grid of twice the problem size in each dimension. The magnetization field $\Mb$ is zero-padded to match this extended domain, enabling a periodic convolution in Fourier space that is equivalent to a linear convolution under open boundary conditions. This approach avoids spurious periodic artifacts and enables accurate evaluation of long-range dipolar interactions.

After each time step or stage evaluation, the magnetization vector $\Mb$ is renormalized at every grid point to enforce $|\Mb| = M_s$, consistent with the physical constraint of constant saturation.

This spatial discretization, in combination with modular treatment of effective field terms, enables compatibility with multirate time integration, discussed next. In particular, the computational cost of the demagnetization step -- dominated by FFTs -- motivates assigning it to a slower time scale relative to the rapidly evolving exchange and DMI components.

\subsection{Time Integration with SUNDIALS}
One novel capability in MagneX is the use of SUNDIALS to control the time-stepping strategy.
The ARKODE package \cite{reynolds2023arkode} within SUNDIALS provides MagneX with an extensive library of explicit, implicit, and implicit-explicit (ImEx) Runge--Kutta methods. 
Additionally, ARKODE includes multirate infinitesimal (MRI) schemes \cite{schlegel2009multirate, sandu2019class, chinomona2021implicit}, where different physical processes can be advanced with different step sizes using a combination of explicit, implicit, or ImEx approaches. Both classes of methods also support adaptive step sizes where the overall time step, $\Delta t$, is varied based on user-defined accuracy tolerances.
The overall goal is to choose settings that balance trade-offs in temporal accuracy, stability, and time-to-solution.

The flexibility of SUNDIALS enables splitting the right-hand side, $\MRb$, of equation (\ref{eq:LLG}) in various ways. We consider the three term partitioning,
\begin{equation}
\MRb = \MRb^E + \MRb^I + \MRb^F.
\end{equation}
The $\MRb^E$ and $\MRb^I$ terms are advanced with explicit and implicit discretizations, respectively, using the full time step and are referred to as the ``slow-explicit'' and ``slow-implicit'' right-hand-side terms.
The $\MRb^F$ partition contains terms that are integrated with a smaller step size than applied to $\MRb^E$ and $\MRb^I$, and is denoted as the ``fast'' right-hand-side term.
To address stability concerns, numerically stiff terms are typically incorporated into $\MRb^I$ and/or $\MRb^F$; the former has the numerical stability advantage of implicit schemes, and the latter reduces the time step used for only that process.
In micromagnetics codes, the limit on the time step is often dominated by stiff exchange physics, particularly as the grid is refined, as the allowable time step scales as the inverse of the square of the grid spacing. While demagnetization is more computationally-expensive, the step size constraint is not as strict.
Leveraging these differences, we demonstrate that the time-to-solution of classical Runge--Kutta approaches can be improved upon with additively partitioned methods.
Note that all three partitions in $\MRb$ do not need to be used; perfectly acceptable combinations (each of which will be demonstrated in the Results section) include only $\MRb^E$ (explicit), $\MRb^E$ and $\MRb^I$ (ImEx), and $\MRb^E$ and $\MRb^F$ (explicit MRI).
Combined with the choices of integration strategies for the slow partitions, MRI methods allow for extreme flexibility in the methods applied to problem components. Previous studies with MRI methods have demonstrated greater computational efficiency compared to single-rate methods in combustion \cite{loffeld2024performance} and non-equilibrium quantum systems \cite{yao2025advancing} where the time scales of physical processes can differ by several orders of magnitude.

\subsection{Machine Learning Module Implementation}
To enable data-driven computation within MagneX, we have developed an NN module that serves as an alternative backend for the computation of the demagnetizing field, $\mathbf{H}_\text{demag}$. This framework is designed to be modular and minimally invasive to the core solver logic, allowing NN-based surrogates to replace the conventional FFT-based demagnetization with no change to the surrounding numerical integration pipeline.

The ML surrogate model is trained using an offline strategy, in which datasets of magnetization field inputs, $\mathbf{M}$, and corresponding demagnetizing field outputs, $\mathbf{H}_\text{demag}$, are generated via forward simulations using the conventional solver. These datasets are used to train neural operators in Python using the PyTorch framework. Once trained, models are exported via \texttt{torch.jit.trace} into TorchScript format. This produces a static computation graph suitable for deployment in a C++ environment without requiring any Python dependencies at runtime. The serialized TorchScript model is then loaded into the MagneX backend using the LibTorch C++ API.

All inference operations, including input transformation and NN evaluation, are executed on the GPU without transferring data to or from the CPU. Magnetization data, $\mathbf{M}$, is stored in AMReX \texttt{MultiFab} structures, which use Fortran-style column-major memory layout. To ensure compatibility with PyTorch’s row-major tensor format, a layout-aware data transformation is implemented. This transformation stacks the vector field components into a 4D tensor of shape $(N_b, C, H, W)$, where $N_b$ is the batch size (typically 1), $C = 3$ channels for $(M_x, M_y, M_z)$, and $(H, W)$ denote the in-plane spatial dimensions. This tensor is passed to the neural model on device, with no host-device transfers required.

\begin{figure*}[!t]
    \centering
    \includegraphics[width=1\textwidth]{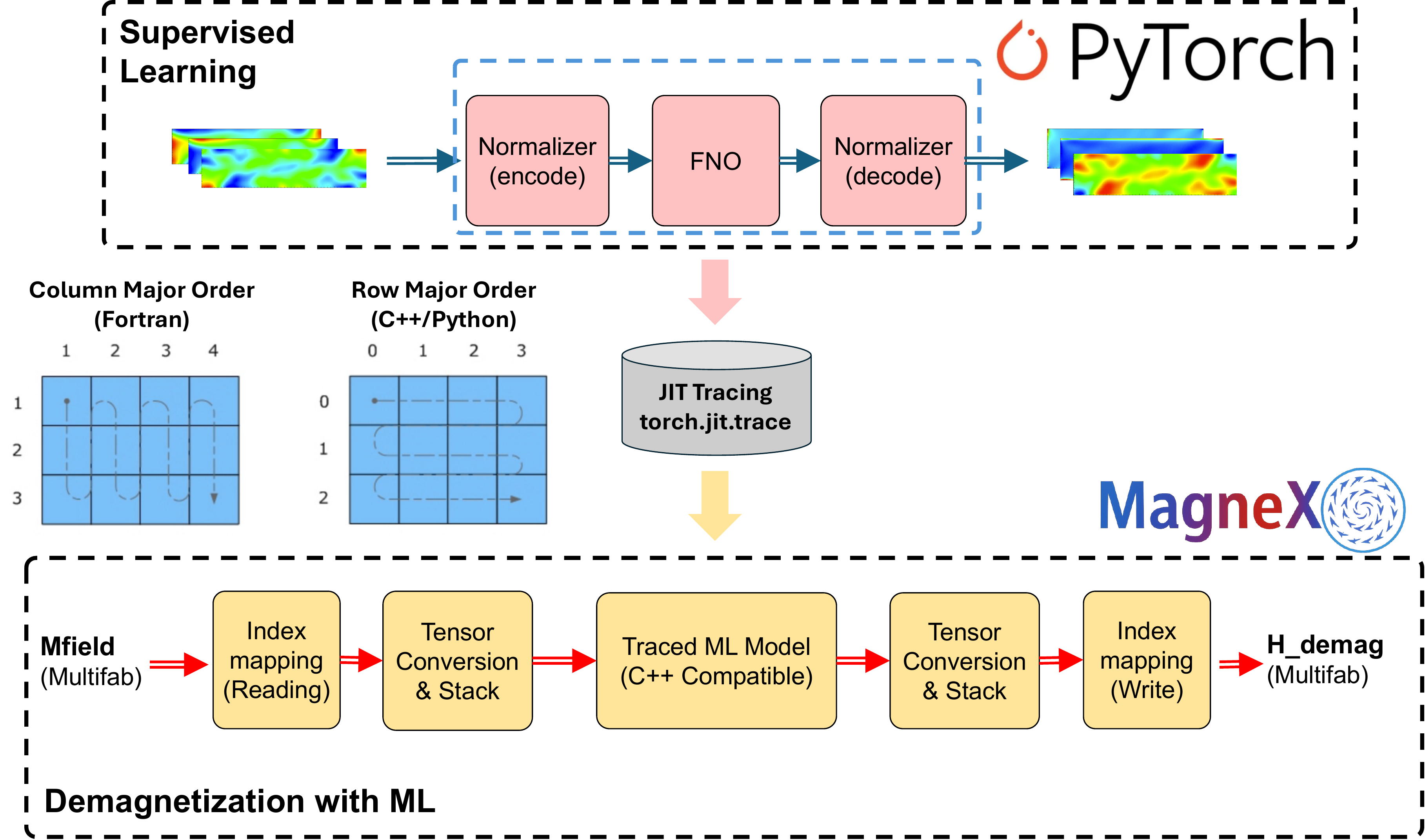}
    \caption{\textbf{Machine learning pipeline for demagnetization in MagneX.} A supervised learning framework based on PyTorch is used to train a Fourier Neural Operator (FNO) model to approximate the demagnetizing field, $\Hdemag$, from the magnetization field, $\mathbf{M}$. The trained model is JIT-traced for C++ compatibility and integrated into the MagneX simulation workflow. The pipeline includes data normalization, tensor conversion between column-major (Fortran) and row-major (C++/Python) memory layouts, and bidirectional index mapping between AMReX’s Multifab data structures and PyTorch tensors. This enables efficient, runtime evaluation of learned surrogates during micromagnetic simulations.}
    \label{fig:ml_pipeline}
\end{figure*}

After inference, the predicted demagnetizing field, $\mathbf{H}_\text{demag}$, is mapped back into the \texttt{MultiFab} data structure using an inverse transformation that restores the correct spatial layout and component indexing. This bidirectional tensor-field mapping ensures numerical consistency and enables seamless substitution of the NN-based demagnetization in the time integration loop. The architecture also supports extensions to batched inference, multi-frame updates, and context-aware hybrid solver logic. Figure~\ref{fig:ml_pipeline} illustrates the overall dataflow and integration strategy.

\subsubsection{Dataset Curation$\And$ML Training details}\label{subsubsec2}

The training dataset for the machine learning model is generated using the conventional MagneX solver configured to simulate a thin-film permalloy system representative of $\mu$MAG Standard Problem 4 (see Section \ref{sec:std4} for full simulation details).
To induce diverse magnetization responses, once the S-state is established, constant external magnetic fields, $\mathbf{H}_\text{bias}(t)$, are sampled randomly, in both magnitude and direction, and applied to the system.
Field magnitudes are selected to fall within a physically relevant range (approximately 25–40 mT), and field directions are sampled uniformly over the unit sphere and projected into the film plane.

For each random field sample,
the system is integrated for 400,000 time steps with a fixed step size of $5 \times 10^{-15}$~s with a Forward Euler method.
Snapshots of the magnetization field, $\mathbf{M}(\mathbf{r}, t)$, and the corresponding demagnetizing field, $\mathbf{H}_\text{demag}(\mathbf{r}, t)$, are stored every 20,000 time steps, resulting in 20 frame pairs per simulation. A total of 1,000 such simulations are performed, yielding 20,000 input-output pairs for supervised training. The data is stored in NumPy array format and subsequently normalized using a per-channel Gaussian normalizer for training.

The neural surrogate model used for demagnetization field prediction in MagneX is based on a two-dimensional Fourier Neural Operator (FNO). The model was trained offline in a supervised fashion to approximate the mapping from normalized magnetization fields, $\mathbf{M} \in \mathbb{R}^{3 \times H \times W}$, to the corresponding demagnetizing field, $\mathbf{H}_\text{demag} \in \mathbb{R}^{3 \times H \times W}$.

All training data is preprocessed using channel-wise normalization. Specifically, a unit Gaussian normalizer was computed across the training set to ensure that each component of $\mathbf{M}$ and $\mathbf{H}_\text{demag}$ had zero mean and unit variance. These transformations were stored and applied consistently to both training and inference datasets to ensure numerical stability. Input and target tensors are stored in $(C, H, W)$ format and reshaped to match the $(N_b, H, W, C)$ convention expected by the FNO model, where $N_b$ is the batch size and $C = 3$ corresponds to the Cartesian components.

The FNO model employed consists of an input lifting layer that projects the 3 input channels to a 32-dimensional latent space, followed by four spectral convolution layers operating in Fourier space. Each spectral block performs a truncated 2D FFT, applies complex-valued learned filters to the first 12 spatial modes in each dimension, and returns to physical space via inverse FFT. A final projection layer reduces the output back to 3 channels corresponding to $(H_x, H_y, H_z)$. No coordinate grids or positional encodings were used, as the domain is spatially uniform and implicitly represented through the spectral basis.

Training was performed using the Adam optimizer with an initial learning rate of $10^{-3}$ and a weight decay coefficient of $10^{-4}$. The learning rate was reduced by a factor of 0.5 at epochs 20, 40, and 60 via a multistep scheduler. The objective function was the mean absolute error (MAE), implemented as a smooth L1 loss,
\[
\mathcal{L}(\theta) = \frac{1}{N} \sum_{i=1}^{N} \text{SmoothL1}\left(f_\theta(\mathbf{M}_i) - \mathbf{H}_i^\text{demag}\right),
\]
where $f_\theta$ is the FNO model with parameters $\theta$ and $N$ is the number of samples per batch.

The model was trained for 100 epochs using a batch size of 128 for training and 32 for validation. Evaluation was conducted after each epoch on a held-out validation set to monitor generalization error. All training and inference were executed on a single NVIDIA A100 GPU.

\section{Numerical Validation}

We validate MagneX through a series of standardized micromagnetic benchmarks established by the Micromagnetic Modeling Activity Group ($\mu$MAG), as well as published vortex and skyrmion test cases. 
Our testing encompasses established standards and advanced dynamical cases, structured as follows:

\begin{itemize}
    \item $\mu$MAG Standard Problems 2 \& 3: Quasi-static magnetization evolution, validating the conventional solver components.
    \item $\mu$MAG Standard Problem 4: Dynamic domain wall motion, used to validate numerical integration and the ML demagnetization surrogate, while also demonstrating accelerated computation via SUNDIALS multirate integrators.
    \item Confined Vortex/Skyrmion Dynamics: Simulation of topologically nontrivial states with DMI~\cite{rohart2013skyrmion}, validating the DMI implementation and benchmarking the ML surrogate's accuracy and generalization against the full FFT-based solver.
\end{itemize}

This multi-faceted validation confirms the solver's accuracy across fundamental and complex micromagnetic scenarios.


\subsection{$\mu$MAG Standard Problem 2: Hysteresis}
\begin{figure}[h]
    \centering
    \includegraphics[width=0.25\textwidth]{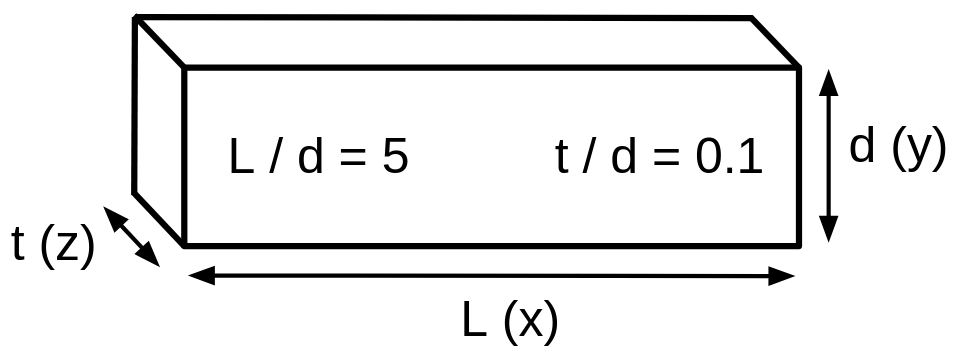}\\
    \includegraphics[width=0.425\textwidth]{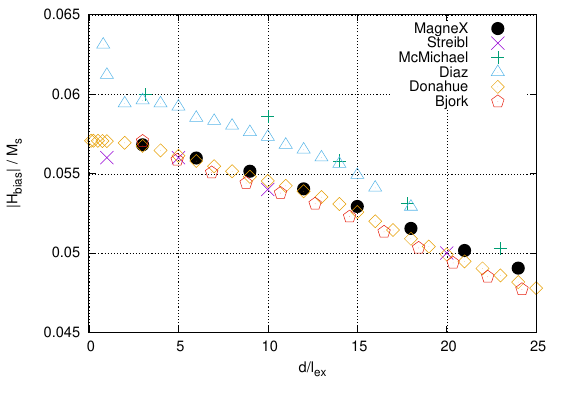}\\
    \includegraphics[width=0.425\textwidth]{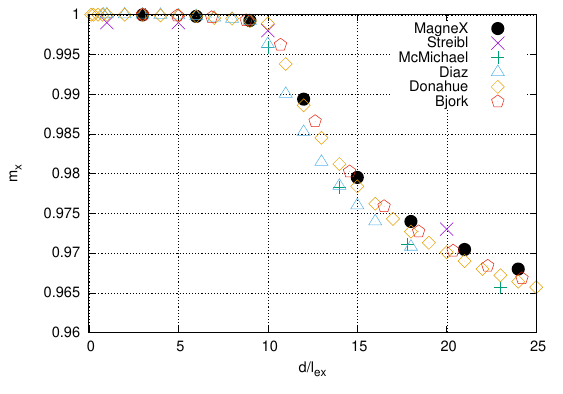}\\
    \includegraphics[width=0.425\textwidth]{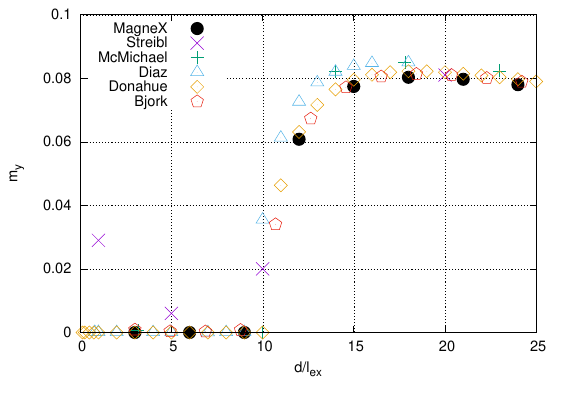}
    \caption{\textbf{Validation of MagneX against $\mu$MAG Standard Problem 2.} 
    (Top) Geometry of the thin ferromagnetic prism used in the benchmark, with aspect ratios $L/d = 5.0$ and $t/d = 0.1$, as specified by the problem. 
    The next three panels show a comparison of MagneX results with reference data for the normalized coercive field, and $m_x$ and $m_y$ remanent magnetization components as functions of normalized size $d/\ell_\mathrm{ex}$. 
    The benchmark examines the transition from coherent rotation to vortex-mediated reversal as $d/\ell_\mathrm{ex}$ increases. 
    MagneX results show strong agreement with previously submitted solutions hosted on the $\mu$MAG website.
    }
    \label{fig:std_2_rem}
\end{figure}

In this problem we examine the hysteresis of a magnetic block of material with demagnetization and exchange physics.
The length in $x$ ($L$), width in $y$ ($d$), and thickness in $z$ ($t$) are in proportion of $L/d=5$ and $t/d=0.1$.
We compute the coercivity and remanence as a function of $d/l_{\rm ex}$, where $l_{\rm ex} = \sqrt{A/K_m}$ is the exchange length and $K_m = 0.5\mu_0 M_s^2$ is the magnetostatic energy density.
Here we use $A = 1.005154 \times 10^{-11}$~J/m, $M_s = 8\times 10^5$~A/m, $\alpha = 0.5$, and thus $l_{\rm ex} = 5$~nm.
We disable the precession term as it does not affect the steady configuration at each increment \cite{bjork2021magtense}.
To complete the magnetic hysteresis, we step the external field from $\Hbias = [0.08M_s,~0.08M_s,~0.08M_s]$ to $\Hbias = [-0.08M_s,-0.08M_s,-0.08M_s]$ in 1,000 increments in each direction of the sweep and equilibrate each step until the normalized average value of each component of $\Mb$ changes by less than $10^{-9}$ over a time step.
This tolerance has been validated by observing that the remanences and coercivities do not significantly change with tighter tolerances.
For each simulation we use $50 \times 10 \times 1$ computational grid cells and test values of $d/l_{\rm ex}$ of 3, 6, 9, 12, 15, 18, 21, and 24 using the classical fourth-order Runge--Kutta method (RK4).
We extract the remanences and coercivities from the hysteresis curves and compare them to published results on the $\mu$MAG website \cite{mumag_std2} in Figure \ref{fig:std_2_rem}.
Our results are consistent with the published results.

\subsection{$\mu$MAG Standard Problem 3: Single Domain Limit}

\begin{figure}[h]
    \centering
        \includegraphics[width=0.35\textwidth]{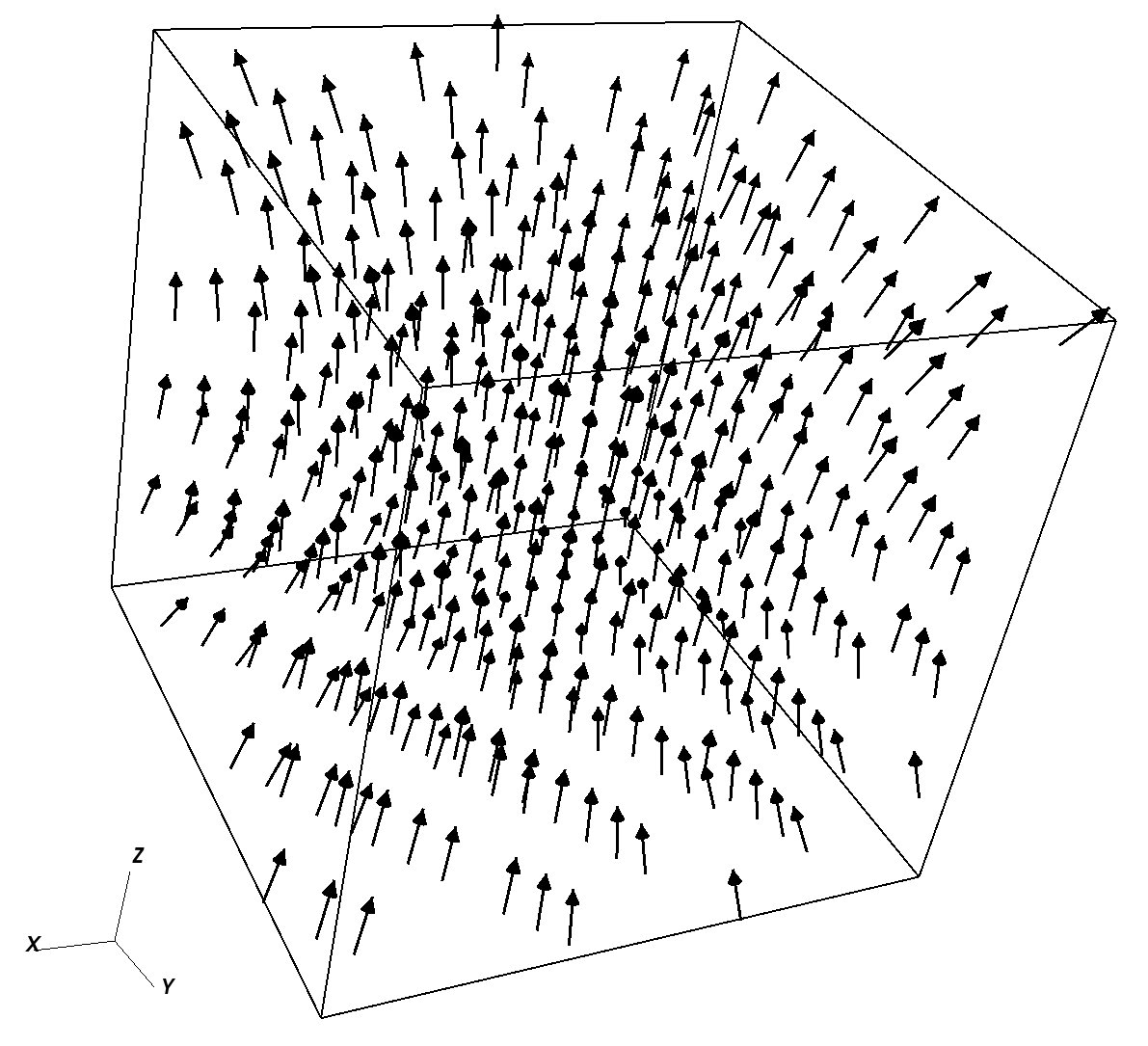}
        \includegraphics[width=0.35\textwidth]{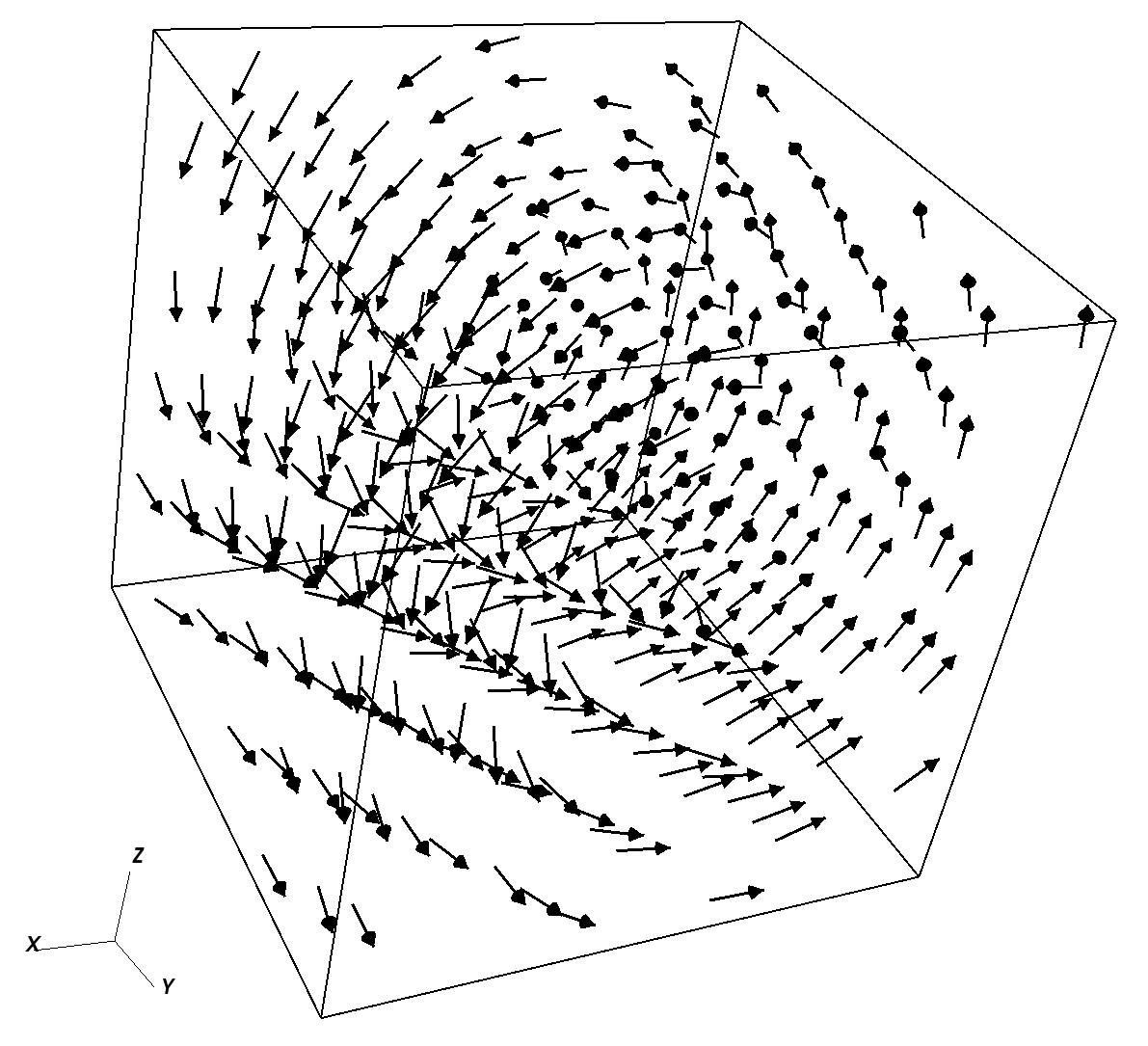}
    \centering
    \caption{$\Mb$ vectors illustrating (top) flower and (bottom) vortex states generated at the critical domain size $L/l_{\rm ex} = 40$.}
    \label{fig:flower_vortex}
\end{figure}

In this problem we consider a cubic block of magnetic material with demagnetization and exchange physics as well as crystalline anisotropy.
The goal is to compute the length of a cube, $L$, expressed in units of $l_{\rm ex}$, such that the energies of two stable states (flower and vortex) are the same.
In this problem we choose $A = 1.005154 \times 10^{-11}$~J/m, $M_s = 8\times 10^5$~A/m, and $\alpha = 0.5$; thus $l_{\rm ex} = 5$~nm.
Uniaxial $\Hani$ is applied with $\eb_K$ pointing along z-axis with a value of $K_u = 0.1K_m = 4.021216\times 10^4$ J/m$^3$.
A stable flower state can be achieved by initializing the magnetization as $\Mb = [M_s/\sqrt{3}, M_s/\sqrt{3}, M_s/\sqrt{3}]$.
Meanwhile, a stable vortex pattern can be achieved by initializing $\Mb = [M_s,0,0]$ in the lower-half of the domain $x$ and $\Mb = [-M_s,0,0]$ in the upper-half of the domain in $x$.
We evolve these fields for 2~ns using RK4, at which point the fields equilibrate to the stable patterns.
In this problem, $N$ is the number of grid cells in each coordinate direction.
For $N=10, 20$, and $40$ we performed simulations to determine the value of $L$ for which the total combined energy (demagnetization, exchange, and anisotropy, as defined in equations: \ref{energy_demag}, \ref{energy_exch}, \ref{energy_ani}) for the flower and vortex states are equal.
We then use Richardson extrapolation to predict $L$ for the infinite resolution $(N\rightarrow\infty)$ case.
In Table \ref{tab:energy} we report our results as well as a comparison to those on the $\mu$MAG website, and see excellent agreement.
In Figure \ref{fig:flower_vortex}, we illustrate the vector fields of the stable states.

\begin{table*}[t]  
\centering
\resizebox{\textwidth}{!}{%
\begin{tabular}{|c|c|c|c|c|c|c|c|c|c|c|}
\hline
\multicolumn{3}{|c|}{ } & \multicolumn{4}{|c|}{Flower} & \multicolumn{4}{c|}{Vortex}\\
\hline
\hline
$N$ & $L$ & $e$ & $e_{\rm demag}$ & $e_{\rm exch}$ & $e_{\rm anis}$ & $m_z$ & $e_{\rm demag}$ & $e_{\rm exch}$ & $e_{\rm anis}$ & $m_y$ \\
\hline
10 & 8.1669 & 0.3038 & 0.2826 & 0.0161 & 0.0051 & 0.9736 & 0.0849 & 0.1663 & 0.0526 & 0.3715\\
\hline
20 & 8.3896 & 0.3030 & 0.2802 & 0.0173 & 0.0055 & 0.9717 & 0.0799 & 0.1708 & 0.0522 & 0.3568\\
\hline
40 & 8.4505 & 0.3028 & 0.2796 & 0.0176 & 0.0056 & 0.9712 & 0.0786 & 0.1719 & 0.0521 & 0.3529\\
\hline
$N\rightarrow\infty$ & 8.4708 & 0.3027 & 0.2794 & 0.0177 & 0.0056 & 0.9710 & 0.0783 & 0.1723 & 0.0521 & 0.3515\\
\hline
Rave et al. & 8.47 & 0.3027 & 0.2794 & 0.0177 & 0.0056 & 0.971 & 0.0783 & 0.1723 & 0.0521 & 0.352\\
\hline
Martins et al. & 8.4687 & 0.3026 & 0.2792 & 0.0177 & 0.0056 & 0.9710 & 0.0780 & 0.1724 & 0.0521 & 0.3516\\
\hline
Bjork et al. & 8.4556 & 0.3029 & 0.2801 & 0.0174 & 0.0055 &  & 0.0785 & 0.1725 & 0.0520 & \\
\hline
\end{tabular}%
}
\caption{Standard Problem 2 test results showing the domain length $L$ (in units of $l_{\rm ex}$) and total and partial energy densities (in units of $K_m$) compared to $\mu$MAG website results.}
\label{tab:energy}
\end{table*}

\subsection{$\mu$MAG Standard Problem 4: Dynamics}\label{sec:std4}

In Standard Problem 4, we model the time evolution of magnetization in a thin $500 \times 125 \times 3.125$~nm film with demagnetization and exchange physics.
We use material parameters including  $M_s = 8\times 10^5$~A/m, $\alpha = 0.02$, and $A=1.3\times 10^{-11}$~J/m.
There are two bias field configurations associated with this problem: 
\begin{itemize}
    \item \textbf{Field 1:} $\Hbias = [-19576, 3422, 0]$~A/m (approximately 170$^\circ$ counterclockwise from positive $x$-axis).
    \item \textbf{Field 2:} $\Hbias = [-28259, -5013, 0]$~A/m (approximately 190$^\circ$ counterclockwise from positive $x$-axis).
\end{itemize}
In each case we first generate the equilibrium S-state before enabling the external field:
\begin{enumerate}
   \item Setting $\alpha = 0.5$ and initializing $\Mb = [M_s, 0, 0]$.
   \item Applying an external field $\Hbias = [10^5, 10^5, 10^5]$~A/m constant for 20ps, then linearly reducing to zero over 10ps.
   \item Relaxing for 1.0 ns after the diagonal external field vanishes.
\end{enumerate}

\begin{figure}[htb]
    \centering
    \includegraphics[width=0.425\textwidth]{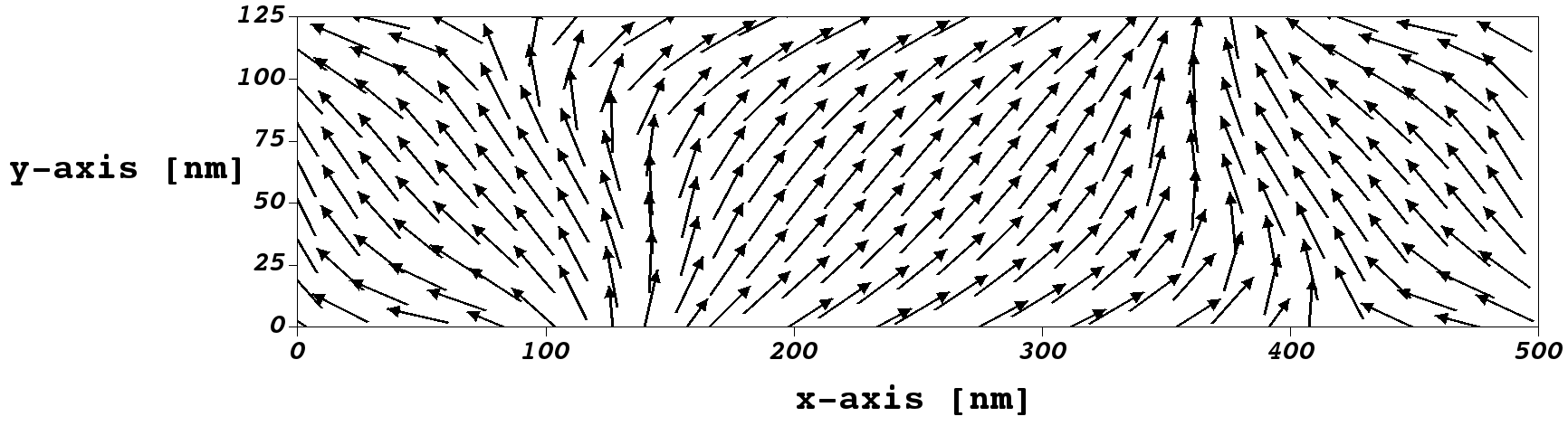}\\
    \includegraphics[width=0.425\textwidth]{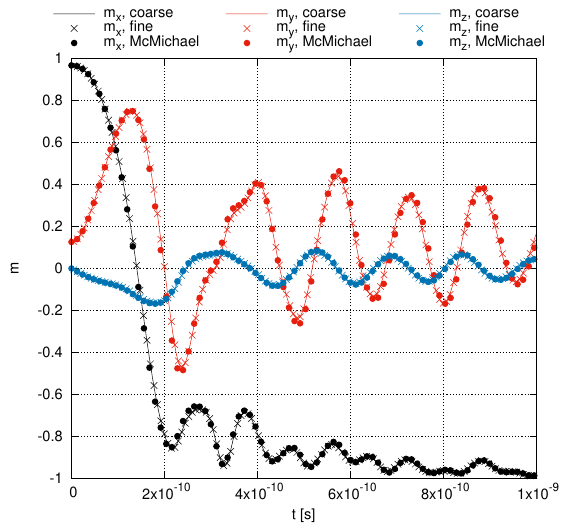}\\    
    \includegraphics[width=0.425\textwidth]{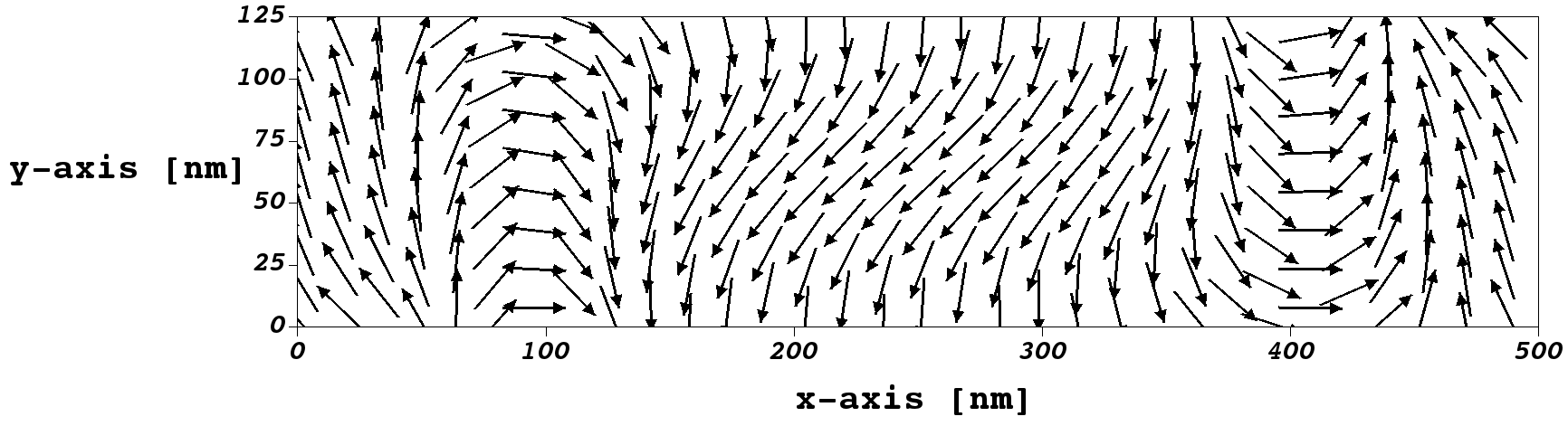}\\
    \includegraphics[width=0.425\textwidth]{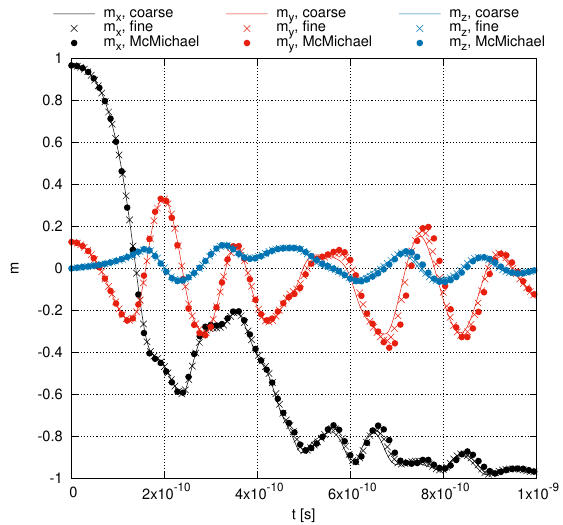}
    \caption{Images of magnetization when $m_x$ first crosses zero and evolution of $(m_x, m_y, m_z)$ for (top) Field 1 and (bottom) Field 2 in Standard Problem 4 at both coarse and fine resolution, and comparison with McMichael et al.~\cite{mcmichael2001switching} $\mu$MAG website results.}
    \label{fig:std4_evolution}
\end{figure}

We use integrate-in-time with the RK4 method. In Figure \ref{fig:std4_evolution}(a) and (b) we present the magnetization vector field when the average value of $M_x$ first crosses from positive to negative for each field.
Our results compare very well to those on the $\mu$MAG website~\cite{mumag_std2}.
For each field we test two different grid resolutions, 3.125~nm and 1.5625~nm.
Figures \ref{fig:std4_evolution}(c) and (d) show excellent agreement between the three  magnetization components at coarse and fine resolutions with the solutions from McMichael et al.~\cite{mcmichael2001switching}.

\subsubsection{Performance of Time Integration with SUNDIALS}

To demonstrate the temporal integrator flexibility provided by SUNDIALS and highlight a use-case where MRI integrators outperform the widely-used RK4 method in time-to-solution, we test different integration strategies and report the performance of each method.
The sheer number of possibilities to consider are beyond the scope of this paper, so here we focus on three distinct approaches. 
First with we consider is the classical RK4 method where all terms remain in the $\MRb^E$ partition, and there are four right-hand-side evaluations per time step.
The second case we consider is an ImEx approach. Motivated by the fact that the exchange physics often limits the overall time step, we partition $\Hexch$ into the slow implicit partition, $\MRb^I$, and all the remaining terms into $\MRb^E$. We use the second-order ImEx method from \cite{Giraldo2013} and solve the implicit systems in each stage with a Newton-Krylov method using unpreconditioned GMRES \cite{Saad:1986} as the linear solver.
The third case we consider is an explicit-explicit MRI approach where we partition $\Hexch$ into the fast partition, $\MRb^F$, and all the remaining terms into $\MRb^E$. For both the fast and slow timescales we use the three-stage, third-order explicit method by Knoth and Wolke \cite{knoth1998implicit}, where the slow step size is the full time step, $\Delta t$, and the fast step size is $0.1\Delta t$.
The MRI algorithm advances the system by integrating the slow timescale processes, $\MRb^E$, with a large time step, while the fast timescale processes, $\MRb^F$, are subcycled between the slow method stages with an additional forcing term to couple with slow timescale.  
This offers two advantages. First, this method remains stable since the exchange physics are integrated over smaller time steps.  Second, the computationally-expensive demagnetization solver is called very few times over any given time interval (compared to the RK4 and ImEx cases we consider), offering computational savings.

In these simulations we use the exact same Standard Problem 4 setup, but with a finer grid spacing of 0.78125~nm to further stiffen the exchange physics compared to the other processes and highlight the potential advantages of MRI schemes.
We have verified that we retain the correct standard problem solution trajectories for each case.
Simulations are performed on a single NVIDIA Quadro GV100 GPU.
\begin{table*}[htb]
    \centering
    \begin{tabular*}{0.65\linewidth}{@{\extracolsep{\fill}}lccc@{}}
\hline
 & RK4 & ImEx & MRI \\
 \hline
Maximum allowable $\Delta t$ [s] & $2.5\times 10^{-14}$ & $5.0\times 10^{-15}$ & $1.25\times 10^{-13}$ \\
Time-to-Solution [s] & 0.133 & 0.920 & 0.069 \\
Exchange evaluations per time step & 4 & 12 (average) & 37 \\
Demagnetization evaluations per time step & 4 & 3 & 3 \\
Time steps & 5 & 25 & 1 \\
Total exchange evals & 20 & 275 & 37 \\
Total demagnetization evals & 20 & 75 & 3 \\
\hline
    \end{tabular*}
    \caption{Runtime and function evaluation statistics over a $1.25\times 10^{-13}$ [s] simulation interval.\label{tab:time_integration}}
\end{table*}

In Table \ref{tab:time_integration} we report on the maximum allowable stable time step through empirical trials, the wallclock time-to-solution over each $1.25\times 10^{-13}$ s simulation interval, the number of exchange and demagnetization evalulations per time step, the number of time steps required to integrate $1.25\times 10^{-13}$ s, and the total number of exchange and demagnetization evaluations over the time interval.

The most prominent result here is the wallclock time-to-solution for the MRI is reduced by 48\% over the next-best option, RK4.  The ImEx scheme is not efficient, presumably due to the lack of preconditioning and the large number of exchange evaluations requires to iterate the implicit system using the GMRES algorithm.
The allowable time step for MRI is a factor of 5 larger than RK4.  So despite the fact that MRI requires 37 exchange evaluations per time interval (compared to 20 for RK4), the MRI approach only requires 3 expensive demagnetization evaluations per time interval (compared to 20 for RK4).  In other words, the fact that RK4 requires 5 times as many time steps to reach the same solution that 1 time step of MRI does leads to far more demagnetization evaluations.  The fact that MRI requires more exchange evaluations does not hurt performance as much since exchange only requires an inexpensive explicit stencil calculation, compared to a pair of expensive Fourier transforms.

\subsubsection{Machine Learning Verification}
To evaluate the effectiveness of the machine learning surrogate in reproducing physically meaningful magnetization dynamics, we performed full time-dependent simulations using the trained FNO model~\cite{li2020fourier} to predict the demagnetizing field, $\mathbf{H}_\text{demag}$, at each time step. The FNO model was integrated into the solver framework as a surrogate for the demagnetizing field computation, replacing the FFT-based module. All other components of the solver, including the LLG time integration and the computation of exchange, anisotropy, and DMI fields, were retained without modification.

\begin{figure}[!t]
    \centering
    \includegraphics[width=0.35\textwidth]{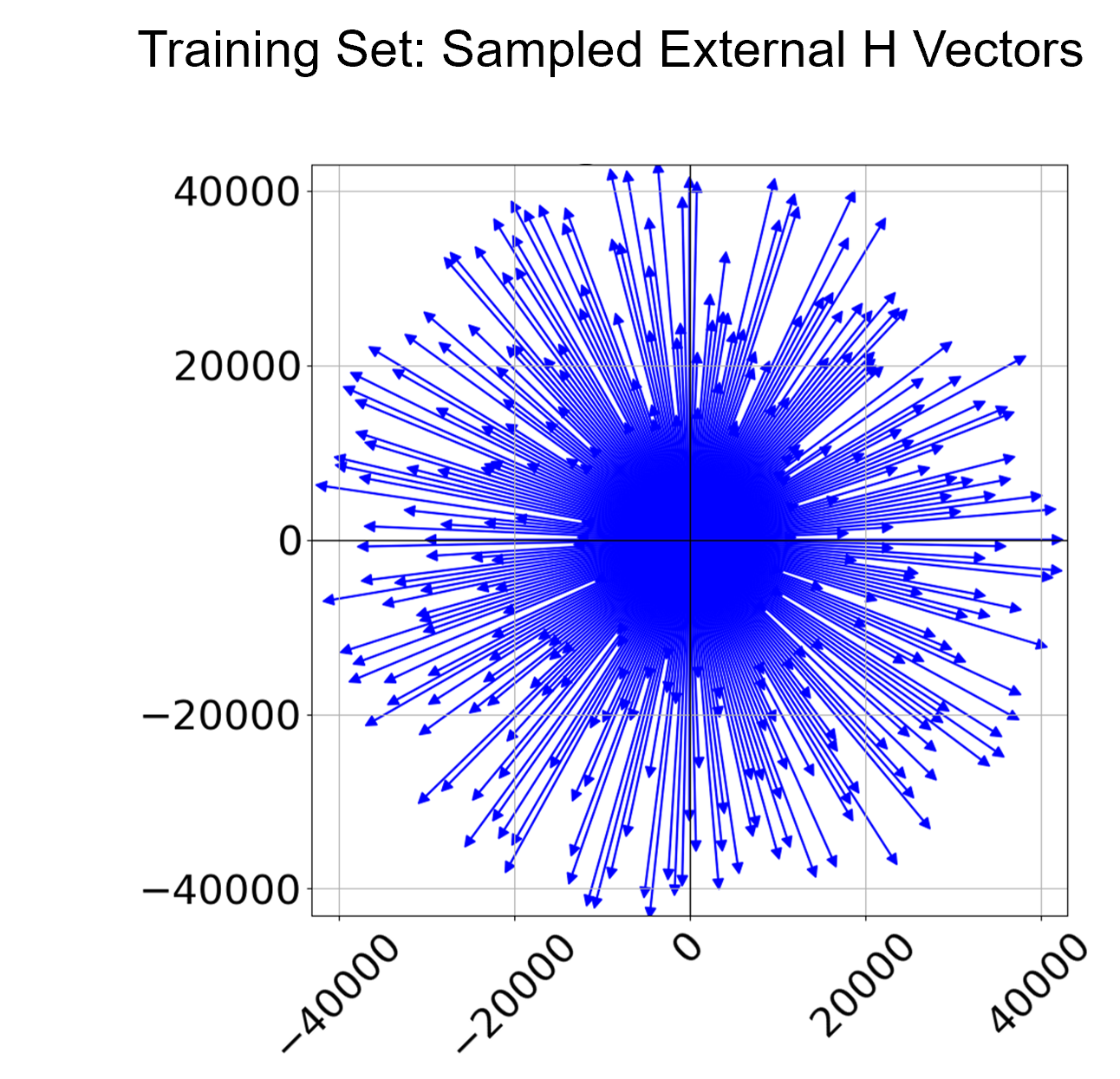}
    \caption{
    Sampling of external magnetic field vectors $\mathbf{H}_{\text{ext}}$ used to generate the training dataset for the ML surrogate.
    }
    \label{fig:std4_fields_a}
\end{figure}

The training data were curated following the strategy described in Section~\ref{subsubsec2}: Dataset Curation \& ML Training details.
Figure~\ref{fig:std4_fields_a} illustrates the sampling distribution of the applied fields. For each simulation, we recorded pairs of magnetization fields $\mathbf{M}$ and their corresponding demagnetizing fields $\mathbf{H}_\text{demag}$, resulting in a dataset suitable for supervised learning.

\begin{figure}[!t]
    \centering
    \includegraphics[width=0.5\textwidth]{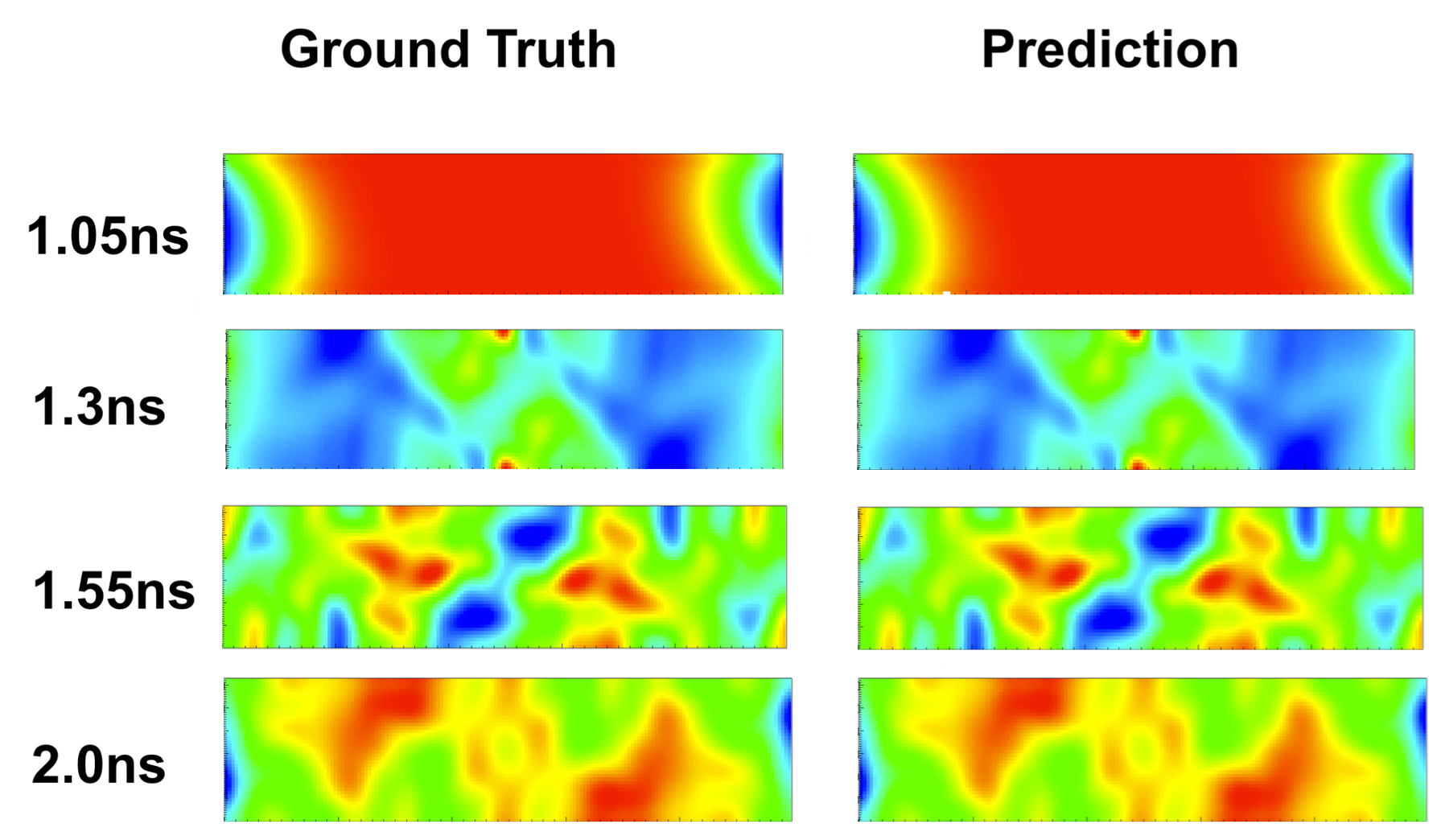}
    \caption{
    Spatial magnetization patterns at selected times for Standard Problem 4 Field 2. The left column shows ground truth results from the FFT-based solver, and the right column shows predictions from the NN-augmented solver with the demagnetizing field replaced by the neural network surrogate.  The colorbar range, from blue to red, in order are $(-1.6\times 10^{4},2.9\times 10^3)$, $(-1.2\times 10^{5},6.1\times 10^5)$, $(-9.9\times 10^{4},1.8\times 10^5)$, and $(-9.1\times 10^{4},9.3\times 10^4)$ [A/m]
    }
    \label{fig:std4_fields_b}
\end{figure}

We now assess the effectiveness of the FNO-based surrogate in reproducing the magnetization dynamics induced by the two prescribed external fields in $\mu$MAG Standard Problem 4. 
To examine spatial fidelity, we present selected snapshots of the $M_z$ component at four time points in Fig.~\ref{fig:std4_fields_b}. The left column shows the ground truth obtained from the original solver, while the right column displays the NN-based predictions at the same time steps. The NN-augmented solver accurately reconstructs the large-scale magnetization features, including domain wall formation and spatial transitions over time. While small-scale differences are present in some fine structures, the overall topological and temporal progression of magnetization is well preserved, indicating that the surrogate model captures the dominant field interactions required to support physically meaningful dynamics.

\begin{figure}[!t]
    \centering
    \includegraphics[width=0.5\textwidth]{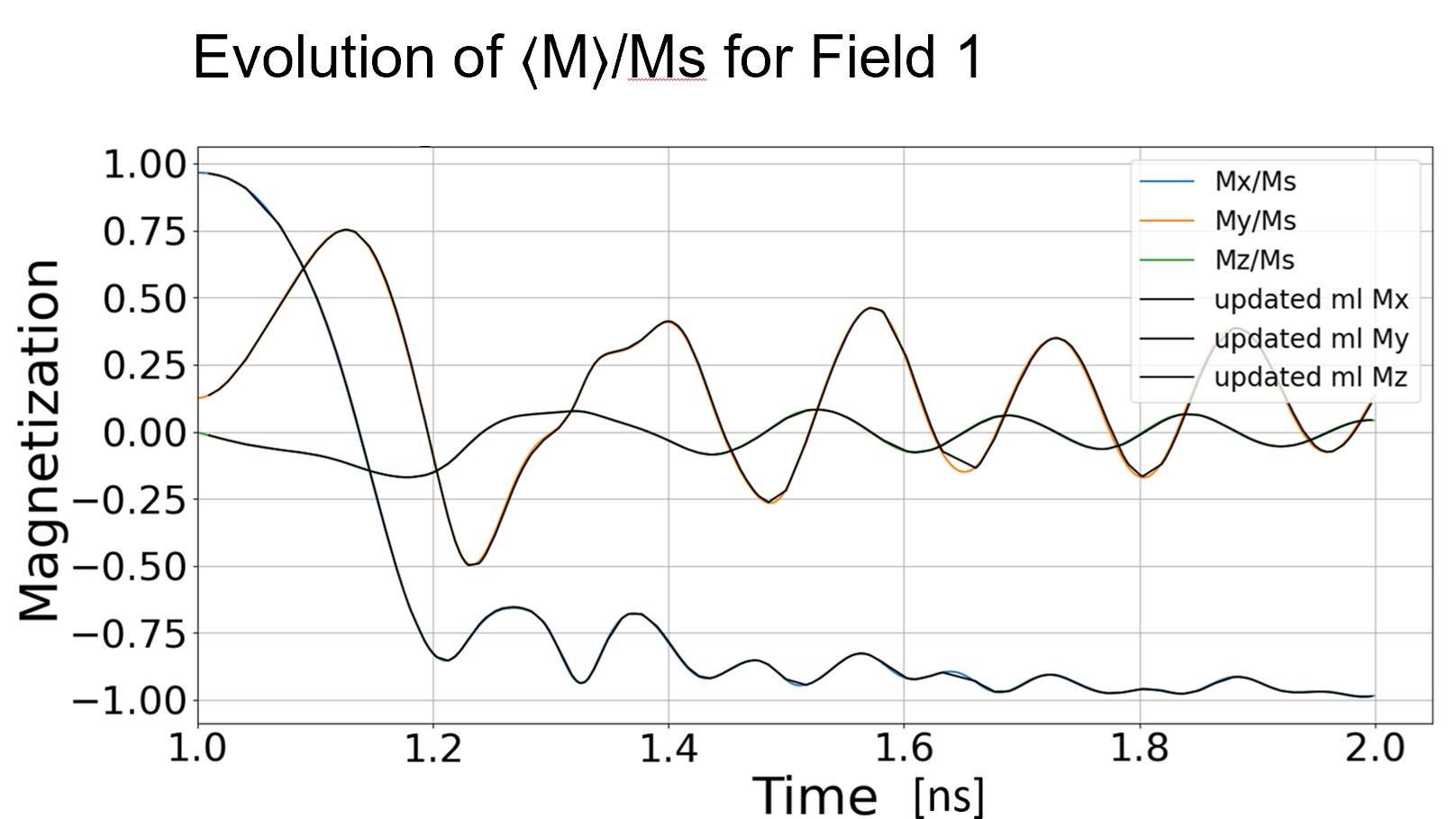}\\
    \includegraphics[width=0.5\textwidth]{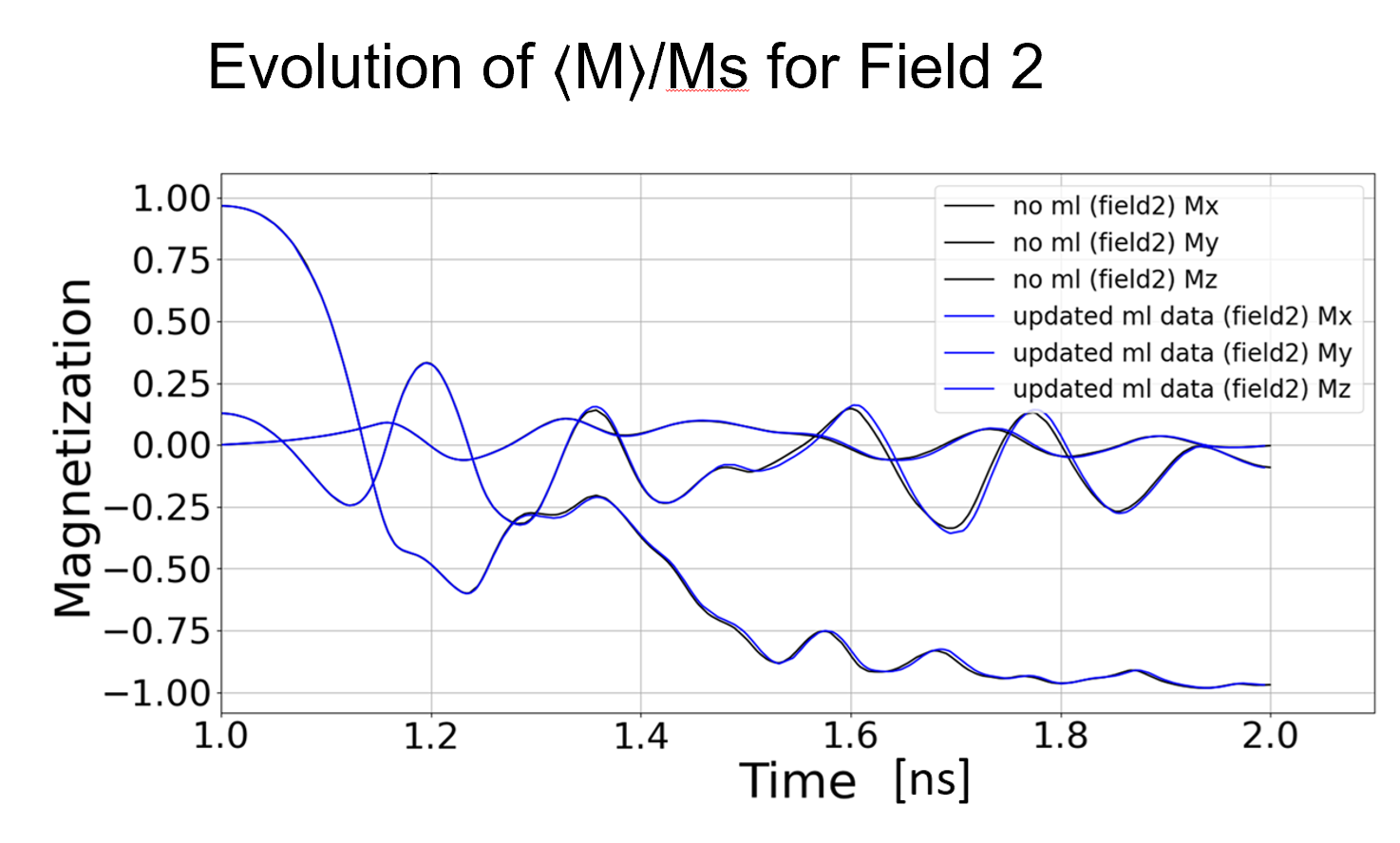}
    \caption{
    Time evolution of the normalized magnetization components under external Fields 1 and 2, comparing results with and without machine learning (ML)-based demagnetization. 
    The close agreement between NN-updated and traditional results demonstrates the viability of data-driven demagnetization in capturing complex dynamic magnetization behavior.
    }
    \label{fig:std4_fields_cd}
\end{figure}

Figure~\ref{fig:std4_fields_cd} shows the evolution of the spatially averaged magnetization components $\langle M_x \rangle$, $\langle M_y \rangle$, and $\langle M_z \rangle$ over time. For both fields, the NN-augmented solver reproduces the global precessional and relaxation trends observed in the conventional solver. In Field 1, the trajectories show strong quantitative agreement across all three vector components. Field 2, which is known to be more sensitive to long-range dipolar interactions and exhibits more complex switching behavior, also shows consistent evolution across the simulation window, with only minor discrepancies.

These results confirm that the neural surrogate provides not only accurate time evolution of global magnetization averages but also physically plausible spatial field structures across a range of dynamical regimes, validating its use as a demagnetization solver in structured micro-magnetic simulations.


\subsection{Vortex (Skyrmion) Verification}
\label{sec:ml_skyrmion}

\begin{figure}[h]
    \centering
    \includegraphics[width=0.4\textwidth]{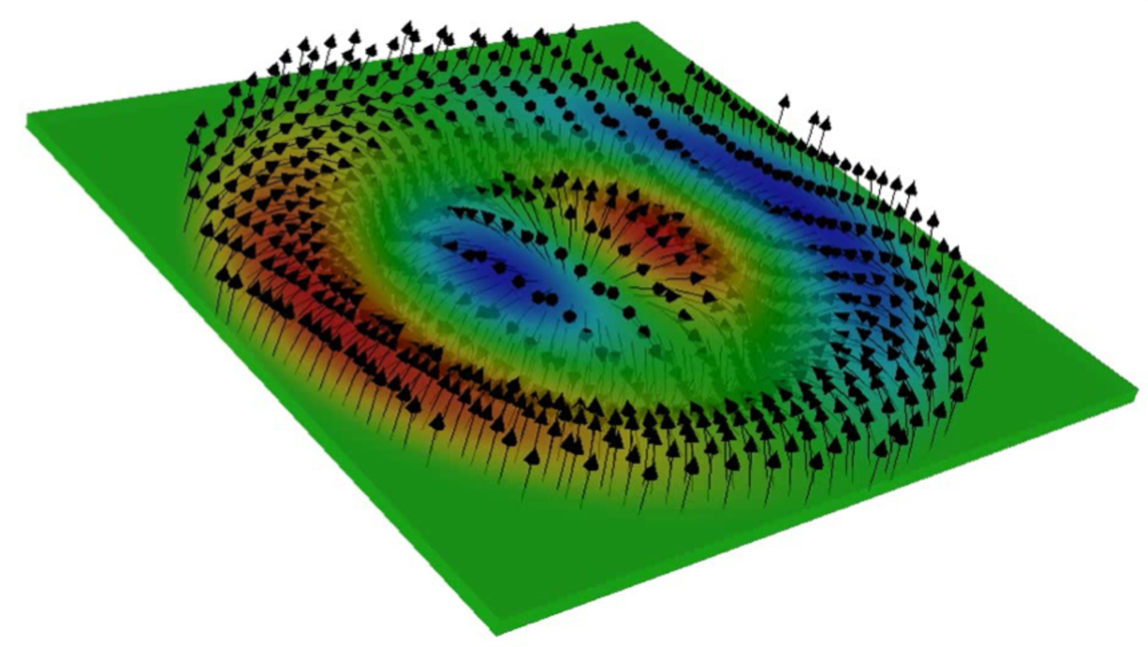}\\
    \includegraphics[width=0.3\textwidth]{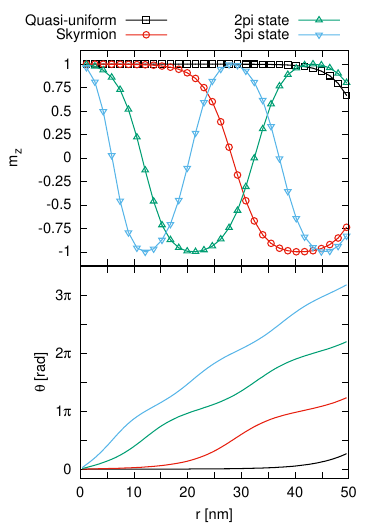}\\
    \includegraphics[width=0.4\textwidth]{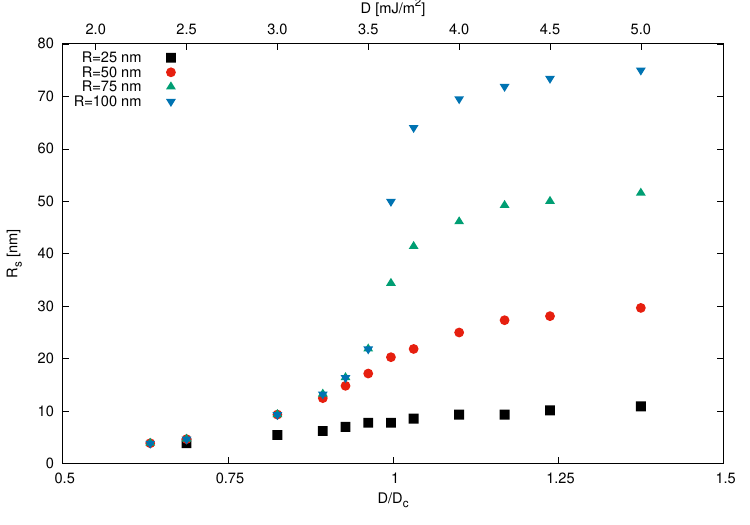}
    \caption{(Top) A vector plot of a steady skyrion profile in a nanodot.
    (Middle) The steady-state simulation results of the $m_z$ profile for a 100-nm-diameter nanodot with $D = 4.5$~[mJ/m$^2$] $(D/D_c = 1.2)$, except for the 3$\pi$ state configuration which requires $D = 5.5$ [mJ/m$^2$].  Also shown is the variation of $\theta$ shows the chirality imposed by DMI in the micromagnetic configuration. For this set of parameters, three solutions are found: quasiuniform (black), skyrmion (red), $2\pi$ (green) and $3\pi$ (blue) rotation states.
    (Bottom) The variation of skyrmion radius $R_s$ as a function of both $D$ and the nanodot radius.
    These results are nearly indistinguishable to the results in Figure 3 of \cite{rohart2013skyrmion}.}
    \label{fig:dmi1}
\end{figure}

We validate our DMI implementation by comparing to stable magnetic configurations in confined geometries as presented in \cite{rohart2013skyrmion}.
The configuration is a nanodot, or a thin cylindrical film of magnetic material.
We use $A = 16\times 10^{-12}$~J/m, $M_s = 1.1\times 10^6$~A/M, $\alpha=0.05$, and $K_u = 5.5\times 10^5$~J/m$^3$.
Demagnetization is not used in these benchmark cases, although we note it has a non-trivial effect on the time-evolution and configuration of the stable states.  Our validation cases use $\Delta x = \Delta y = 0.78125$~nm and $\Delta z = 0.25$~nm.

\subsubsection{Numerical Modeling Verification}
In the first set of tests, we model the time-evolution of magnetization in a $R = 50$~nm radius nanodot until a steady low-energy solution is achieved.
Different solutions are obtained by supplying different initial conditions which are close to each low-energy state.
We obtain the same four solutions as in \cite{rohart2013skyrmion} using $D=4.5$~mJ/m$^2$, namely the quasi-uniform, skyrmion, 2$\pi$, and 3$\pi$ states.  There is one notable exception as our simulations could not reach a stable 3$\pi$ state unless we increased $D$ to $5.5$~mJ/m$^2$.
Figure \ref{fig:dmi1} illustrates a vector plot of the three-dimensional skyrmion profile, as well as the $m_z$ and $\theta$ profiles as a function of radius, where $\theta$ is the angle of the magnetization with respect to the $z$-axis.

In the second set of tests, we a run suites of simulations sweeping $D$ over the range $0.6 < D/D_c < 1.4$ where the critical DMI constant is $D_c = 4\sqrt{AK}/\pi = 3.6371\times 10^{-3}$ and holding the radius of the nanodot constant at $R = 25, 50, 75$, and 100~nm.
In each simulation, we obtain the stable skyrmion state and compute the radius, $R_s$ that defines the skyrmion (i.e., where $m_z$ crosses zero).
In Figure \ref{fig:dmi1} we report the variation of $R_s$ as a function of both $D$ and the nanodot radius.
Our report results are nearly indistinguishable from the results in Figure 4 of \cite{rohart2013skyrmion}.

\subsubsection{Machine Learning Verification}
To further evaluate the generalization capability of the trained NN model beyond standard problem benchmark settings, we applied it to simulate skyrmion-like vortex dynamics in a confined nanodot geometry. These systems exhibit topologically nontrivial magnetization textures, where the balance between exchange, DMI, demagnetization, and anisotropy governs the stability and evolution of the skyrmion core. Unlike the Standard Problem 4 cases, this configuration presents a higher degree of spatial localization and symmetry, posing a distinct challenge for data-driven surrogates.

\begin{figure}[!t]
    \centering
    \includegraphics[width=0.35\textwidth]{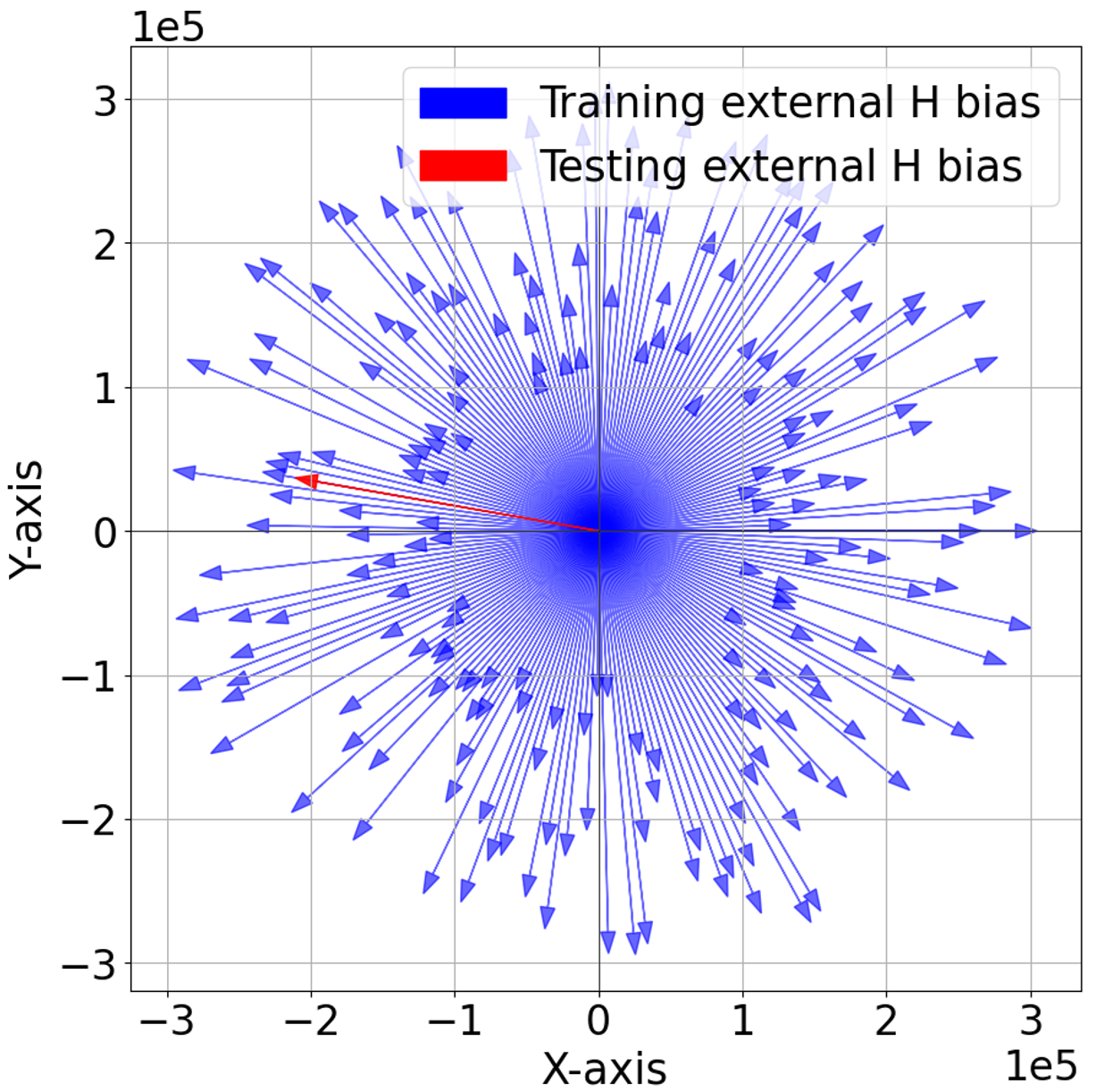}\\
    \includegraphics[width=0.4\textwidth]{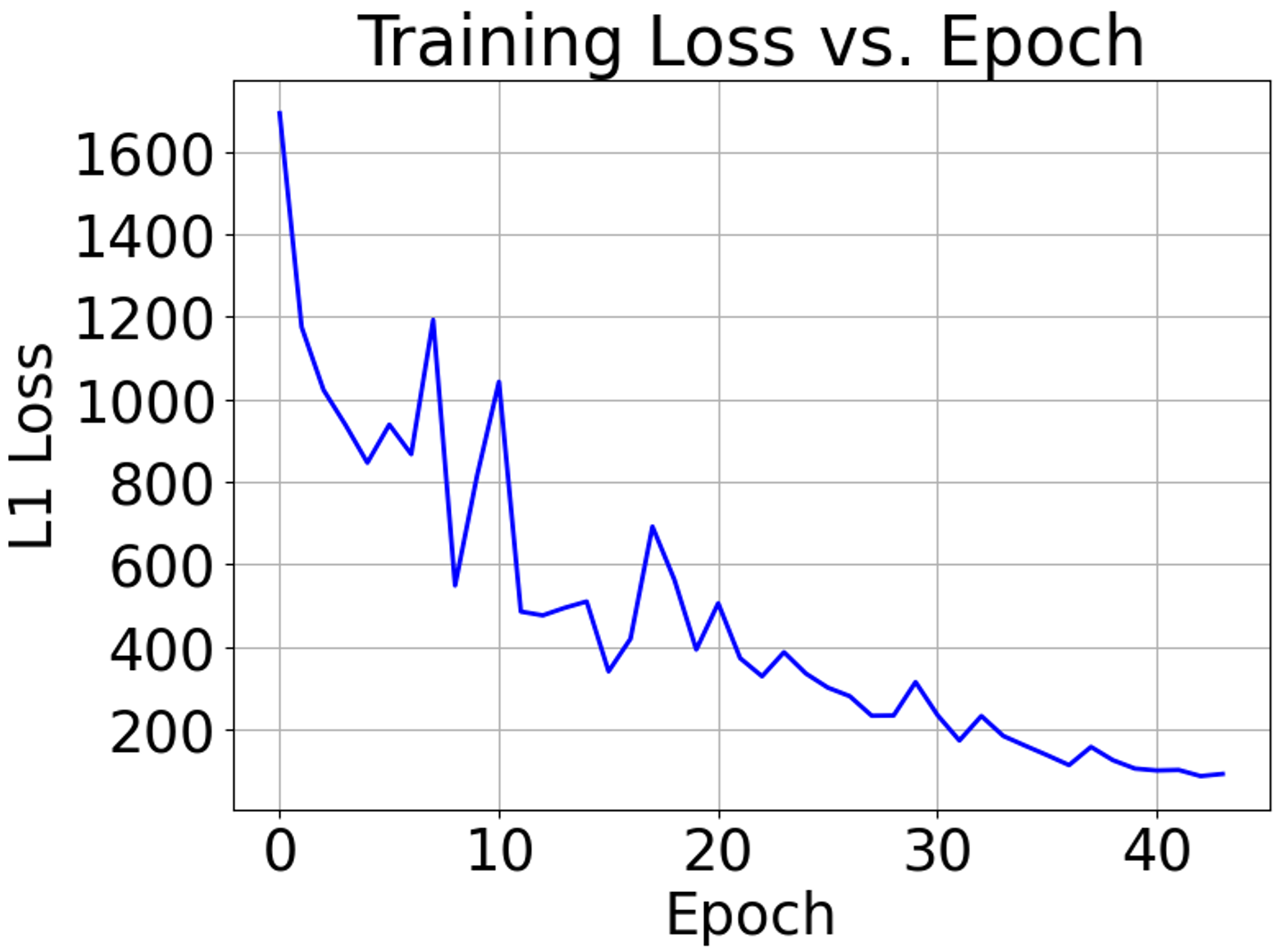}\\
    \includegraphics[width=0.5\textwidth]{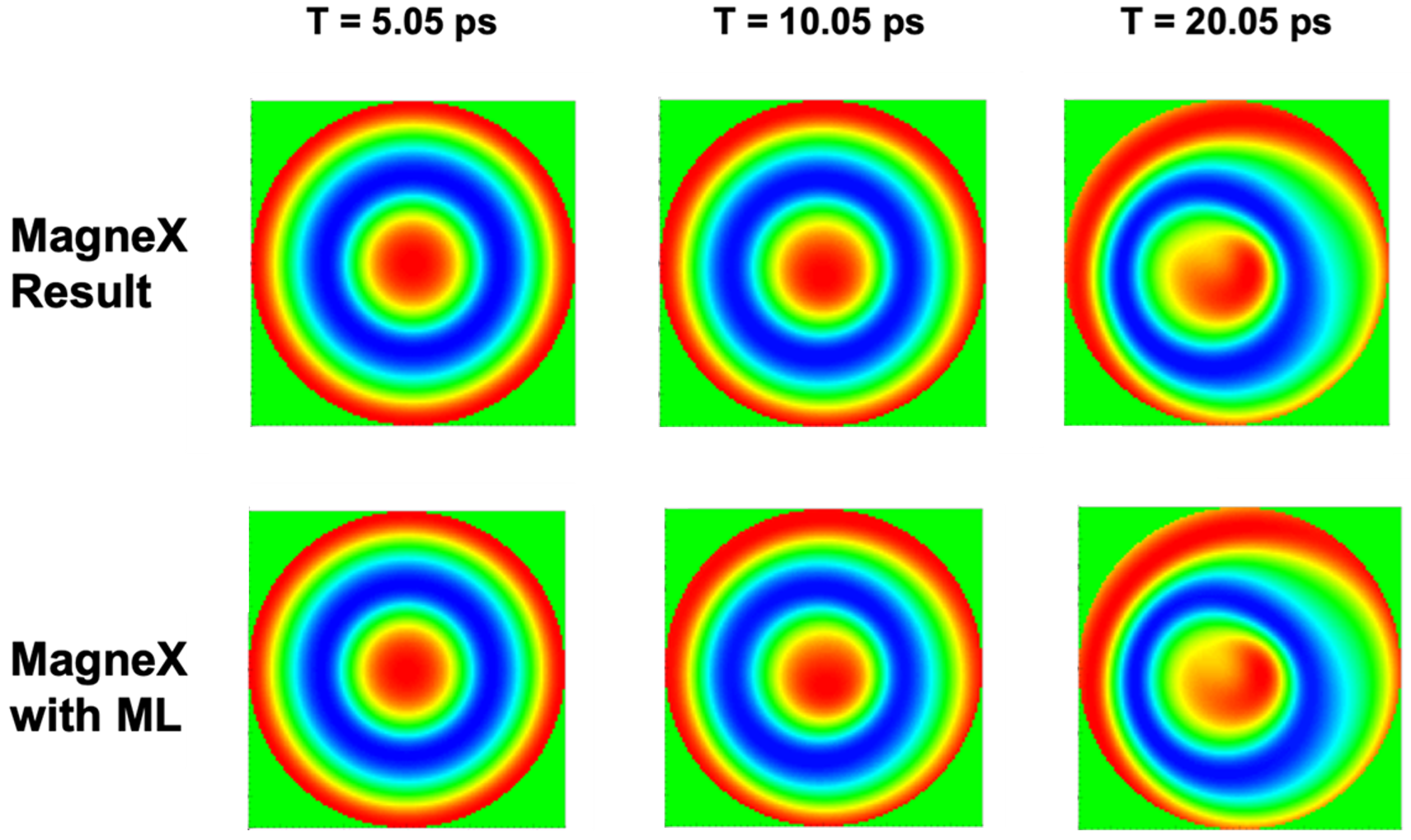}\\
    \caption{\textbf{NN-based demagnetization modeling for vortex evolution in a magnetic nanodot.} 
    (Top) Sampling strategy for generating the training dataset: external magnetic field vectors are sampled uniformly across a range of magnitudes and directions, with red indicating held-out test points. 
    (Middle) Complete training log of the neural network surrogate using L1 loss over epochs. 
    (Bottom) Comparison between the ground truth MagneX simulations and NN-augmented predictions of vortex dynamics at three time points. 
    }
    \label{fig:skyrmion_results}
\end{figure}

Figure~\ref{fig:skyrmion_results} shows the distribution of external magnetic bias fields used for training and testing. The training set consists of 1,000 simulations with randomly sampled field directions and magnitudes, while the test case corresponds to a held-out field vector (highlighted in red). The FNO model was trained offline using L1 loss, with convergence also shown in the figure.

To validate the surrogate, we performed inference using the NN-augmented solver on the test field and compared the results against ground truth MagneX simulations. Figure~\ref{fig:skyrmion_results} displays snapshots of the out-of-plane magnetization component $M_z$ at three representative times (5.05, 10.05, and 20.05 ps). The FNO model successfully captures the concentric ring structure characteristic of vortex/skyrmion states, as well as the radial expansion and phase rotation over time. The agreement is particularly strong in the spatial topology and symmetry of the magnetization profile.


These results demonstrate that the FNO surrogate not only performs well on structured benchmark problems but also generalizes effectively to nontrivial, topologically constrained dynamics. This supports the viability of using NN-based demagnetization surrogates for efficient and physically accurate simulation of complex spin textures in confined magnetic systems.

\subsection{Computation Performance benchmark}
MagneX written in C++ and built on the portable, open-source AMReX library which provides data structures and abstraction layers for computations with MPI, OpenMP, and GPU architectures (NVIDIA, AMD, and Intel) in massively parallel applications on block structured grids.
FFT support is provided through the AMReX-FFT backend; the MagneX implementation presently supports FFTW, cuFFT, and rocFFT backends for pure-MPI, MPI+CUDA (NVIDIA), and MPI+ROCm (AMD) calculations.
Support for SUNDIALS is provided through the AMReX-SUNDIALS interoperability layer that enables SUNDIALS integrators to operate directly on AMReX data structures. A Python-syntax parser is used to define space and time-dependent external fields, material properties, and initial conditions.
All the codes are publicly available on the open-source GitHub platform.
MagneX can be obtained from \wraptt{github.com/AMReX-Microelectronics/MagneX}, AMReX from \wraptt{github.com/AMReX-Codes/amrex}, and SUNDIALS from \wraptt{github.com/LLNL/sundials}.

\begin{figure}[h]
    \centering
    \includegraphics[width=0.45\textwidth]{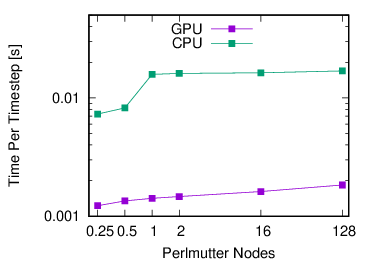}
    \includegraphics[width=0.45\textwidth]{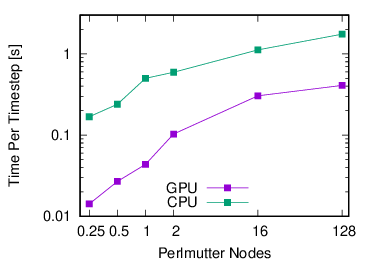}
    \caption{Weak scaling results indicate wallclock time per time step for the cases described in Table \ref{tab:scaling_setup}.
    On the top, all physics except for demagnetization are enabled, whereas on the bottom, all physics are enabled.}
    \label{fig:scaling}
\end{figure}

We perform CPU and GPU weak scaling studies on the NERSC Perlmutter system using two test cases. 
In both setups, the base configuration is a cubic domain with $128^3$ grid cells and a uniform grid resolution of 4~nm.
We use forward Euler for all tests as it is representative of the computational patter in more complex explicit methods (e.g., higher-order Runge--Kutta or explicit-explicit MRI integrators). 
A Perlmutter CPU-node consists of two AMD EPYC 7763 processors, each with 64 physical cores for a total of 128 physical cores while a Perlmutter GPU node contains 4 NVIDIA A100 cards. 
Thus, the base case with $128^3$ cells utilizes 1/4 of a node, comparing the performance of 32 physical CPU cores against 1 GPU.
For the CPU tests we assign an MPI rank and a $32\times 32\times 64$ grid to a CPU core and for the GPU tests we assign an MPI rank and a $128^3$ grid to a GPU.
Table \ref{tab:scaling_setup} shows the number of nodes, number of computational cells, and number of grids used in our weak scaling tests.
In the first weak scaling test, we include all physics in $\Heff$, including exchange, anisotropy, DMI, but exclude the computationally demanding demagnetization.
In the second weak scaling test, we also include demagnetization.
The scaling results are shown in Figure \ref{fig:scaling}.  For the cases without demagnetization the scaling is relatively flat, particularly once the 1-node threashold is reached.
For the cases with demagnetization the scaling is not as ideal but still reasonable considering the communication expense associated with parallel FFTs.
\begin{table*}[htb]
    \centering
    \begin{tabular*}{0.65\linewidth}{@{\extracolsep{\fill}}cccc@{}}
\hline
 \# of nodes & \# of computational cells & \# of grids (GPU case) & \# of grids (CPU case) \\ 
\hline
 1/4 & $128\times 128\times 128$ & 1 & 32\\
 1/2 & $256\times 128\times 128$ & 2 & 64\\
 1 & $256\times 256\times 128$ & 4 & 128\\
 2 & $256\times 256\times 256$ & 8 & 256\\
 16 & $512\times 512\times 512$ & 64 & 2048\\
 128 & $1024\times 1024\times 1024$ & 512 & 16384\\
  \hline
    \end{tabular*}
    \caption{
The total problem size for weak scaling tests, expressed in number of grid cells, for each set of simulations.
For perlmutter GPU-node runs, there are 4 GPUs per node and each MPI rank is assigned to a $128\times 128\times 128$ grid to maintain ideal load balancing.
For perlmutter CPU-node runs, there are 128 cores per node and each MPI rank is assigned to a $32\times 32\times 64$ grid to maintain ideal load balancing.\label{tab:scaling_setup}}
\end{table*}

\section{Conclusions}
We introduced MagneX, an open-source micromagnetics solver built on AMReX that targets modern HPC systems through GPU acceleration, multi-GPU parallelism, and modular effective-field physics. MagneX was validated against $\mu$MAG Standard Problems 2-4 and established DMI/skyrmion benchmarks, demonstrating accurate quasi-static and dynamic behavior across multiple coupling mechanisms. Using SUNDIALS time integrators, we showed that multirate MRI methods can reduce time-to-solution by limiting expensive demagnetization evaluations (48\% faster than RK4 in our representative stiff test), while maintaining correct solution trajectories. We also demonstrated a data-driven demagnetization backend using a Fourier Neural Operator deployed via TorchScript/LibTorch for GPU-resident inference, reproducing both global magnetization trends and key spatial textures. 
Future work will focus on improving implicit-solver performance (e.g., preconditioning) and expanding the accuracy and generality of ML surrogates for broader geometries and multiphysics spintronic applications.

\section*{Acknowledgements}
Y.T., and Z.Y.~were supported by the U.S. Department of Energy, Office of Science, Advanced Scientific Computing Research Program, EXPRESS: 2025 Exploratory Research for Extreme-Scale Science, Analog Compute-in-Memory with Trainable Nonlinear Devices, under contract No. DE-AC02-05-CH11231.
A.N., W.Z., and D.J.G~were supported by the U.S.~Department of Energy, Office of Science, Office of Advanced Scientific Computing Research, Scientific Discovery through Advanced Computing (SciDAC) and Next-Generation Scientific Software Technologies Programs through the FASTMath Institute, under Contract No.~DE-AC02-05CH11231 (LBNL) and No.~DE-AC52-07NA27344 (LLNL). This work was partially performed under the auspices of the U.S. Department of Energy by Lawrence Livermore National Laboratory under Contract DE-AC52-07NA27344. LLNL-JRNL-2014548.
This work was also supported in part by previous breakthroughs obtained through the Laboratory Directed Research and Development (LDRD) Program of Lawrence Berkeley National Laboratory under U.S. Department of Energy Contract No. DE-AC02-05CH11231.
Z.Y., J.C.L., P.K., J.A.M., and C.F.S.~were supported by the U.S.~Department of Energy, Office of Science, the Microelectronics Co-Design Research Program, under contract No.~DE-AC02-05-CH11231 (Codesign of Ultra-Low-Voltage Beyond CMOS Microelectronics) for the development of design tools for low-power microelectronics.
This research used resources of the National Energy Research Scientific Computing Center (NERSC), a DOE Office of Science User Facility supported by the Office of Science of the U.S.~Department of Energy under Contract No.~DE-AC02-05CH11231 and under NERSC GenAI award under No.~DDR-ERCAP0030541. 






\ifCLASSOPTIONcaptionsoff
  \newpage
\fi

\bibliographystyle{IEEEtran}
\bibliography{MagneX}

@misc{mumag_std2,
  title = {{$\mu$MAG}: Micromagnetic Modeling Activity Group},
  author = {{Micromagnetic Modeling Activity Group}},
  howpublished = {\url{https://www.ctcms.nist.gov/mumag/mumag.org.html}},
  year = {1997--2000},
}

@article{bjork2021magtense,
  title={MagTense: A micromagnetic framework using the analytical demagnetization tensor},
  author={Bj{\o}rk, Rasmus and Poulsen, Emil Blaabjerg and Nielsen, Kaspar Kirstein and Insinga, Andrea Roberto},
  journal={Journal of Magnetism and Magnetic Materials},
  volume={535},
  pages={168057},
  year={2021},
  publisher={Elsevier}
}

@article{chang2011fastmag,
  title={FastMag: Fast micromagnetic simulator for complex magnetic structures},
  author={Chang, R and Li, S and Lubarda, MV and Livshitz, B and Lomakin, V},
  journal={Journal of Applied Physics},
  volume={109},
  number={7},
  year={2011},
  publisher={AIP Publishing}
}

@article{d2004nonlinear,
  title={Nonlinear magnetization dynamics in thin-films and nanoparticles},
  author={d’Aquino, Massimiliano},
  journal={PhD, University of Napoles},
  year={2004}
}

@misc{donahue1999oommf,
  title={OOMMF User's Guide: Version 1.0},
  author={Donahue, Michael Joseph and Porter, Donald G},
  year={1999},
  publisher={National Institute of Standards and Technology Gaithersburg, MD}
}

@article{hayashi1996calculation,
  title={Calculation of demagnetizing field distribution based on fast Fourier transform of convolution},
  author={Hayashi, Nobuo and Saito, Koji and Nakatani, Yoshinobu},
  journal={Japanese journal of applied physics},
  volume={35},
  number={12R},
  pages={6065},
  year={1996},
  publisher={IOP Publishing}
}

@article{Giraldo2013,
author = {Giraldo, F. X. and Kelly, J. F. and Constantinescu, E. M.},
title = {Implicit-Explicit Formulations of a Three-Dimensional Nonhydrostatic Unified Model of the Atmosphere (NUMA)},
journal = {SIAM Journal on Scientific Computing},
volume = {35},
number = {5},
pages = {B1162-B1194},
year = {2013},
doi = {10.1137/120876034},
}

@article{Saad:1986,
author = {Saad, Youcef and Schultz, Martin H.},
title = {GMRES: A Generalized Minimal Residual Algorithm for Solving Nonsymmetric Linear Systems},
journal = {SIAM Journal on Scientific and Statistical Computing},
volume = {7},
number = {3},
pages = {856-869},
year = {1986},
doi = {10.1137/0907058}
}

@article{knoth1998implicit,
  title={Implicit-explicit Runge-Kutta methods for computing atmospheric reactive flows},
  author={Knoth, Oswald and Wolke, Ralf},
  journal={Applied numerical mathematics},
  volume={28},
  number={2-4},
  pages={327--341},
  year={1998},
  publisher={Elsevier}
}

@article{kraft2024parallel,
  title={Parallel-in-time integration of the Landau--Lifshitz--Gilbert equation with the parallel full approximation scheme in space and time},
  author={Kraft, Robert and Koraltan, Sabri and Gattringer, Markus and Bruckner, Florian and Suess, Dieter and Abert, Claas},
  journal={Journal of Magnetism and Magnetic Materials},
  volume={597},
  pages={171998},
  year={2024},
  publisher={Elsevier}
}

@incollection{landau1992theory,
  title={On the theory of the dispersion of magnetic permeability in ferromagnetic bodies},
  author={Landau, LALE and Lifshitz, Evgeny},
  booktitle={Perspectives in Theoretical Physics},
  pages={51--65},
  year={1992},
  publisher={Elsevier}
}

@article{lepadatu2023accelerating,
  title={Accelerating micromagnetic and atomistic simulations using multiple GPUs},
  author={Lepadatu, Serban},
  journal={Journal of Applied Physics},
  volume={134},
  number={16},
  year={2023},
  publisher={AIP Publishing}
}

@article{loffeld2024performance,
  title={Performance of explicit and IMEX MRI multirate methods on complex reactive flow problems within modern parallel adaptive structured grid frameworks},
  author={Loffeld, John J and Nonaka, Andy and Reynolds, Daniel R and Gardner, David J and Woodward, Carol S},
  journal={The International Journal of High Performance Computing Applications},
  pages={10943420241227914},
  year={2024},
  publisher={SAGE Publications Sage UK: London, England}
}

@article{mcmichael2001switching,
  title={Switching dynamics and critical behavior of standard problem No. 4},
  author={McMichael, RD and Donahue, Michael J and Porter, Donald G and Eicke, Jason},
  journal={Journal of Applied Physics},
  volume={89},
  number={11},
  pages={7603--7605},
  year={2001},
  publisher={American Institute of Physics}
}

@article{nakatani1989direct,
  title={Direct solution of the Landau-Lifshitz-Gilbert equation for micromagnetics},
  author={Nakatani, Yoshinobu and Uesaka, Yasutaro and Hayashi, Nobuo},
  journal={Japanese Journal of Applied Physics},
  volume={28},
  number={12R},
  pages={2485},
  year={1989},
  publisher={IOP Publishing}
}

@article{reynolds2023arkode,
  title={ARKODE: A flexible IVP solver infrastructure for one-step methods},
  author={Reynolds, Daniel R and Gardner, David J and Woodward, Carol S and Chinomona, Rujeko},
  journal={ACM Transactions on Mathematical Software},
  volume={49},
  number={2},
  pages={1--26},
  year={2023},
  publisher={ACM New York, NY}
}

@article{rohart2013skyrmion,
  title={Skyrmion confinement in ultrathin film nanostructures in the presence of Dzyaloshinskii-Moriya interaction},
  author={Rohart, S and Thiaville, A},
  journal={Physical Review B},
  volume={88},
  number={18},
  pages={184422},
  year={2013},
  publisher={APS}
}

@article{schlegel2009multirate,
  title={Multirate Runge--Kutta schemes for advection equations},
  author={Schlegel, Martin and Knoth, Oswald and Arnold, Martin and Wolke, Ralf},
  journal={Journal of Computational and Applied Mathematics},
  volume={226},
  number={2},
  pages={345--357},
  year={2009},
  publisher={Elsevier}
}

@article{vansteenkiste2014design,
  title={The design and verification of MuMax3},
  author={Vansteenkiste, Arne and Leliaert, Jonathan and Dvornik, Mykola and Helsen, Mathias and Garcia-Sanchez, Felipe and Van Waeyenberge, Bartel},
  journal={AIP advances},
  volume={4},
  number={10},
  year={2014},
  publisher={AIP Publishing}
}

@article{vansteenkiste2011mumax,
  title={MuMax: A new high-performance micromagnetic simulation tool},
  author={Vansteenkiste, Arne and Van de Wiele, Ben},
  journal={Journal of Magnetism and Magnetic Materials},
  volume={323},
  number={21},
  pages={2585--2591},
  year={2011},
  publisher={Elsevier}
}

@article{lakshmanan2011fascinating,
  title={The fascinating world of the Landau--Lifshitz--Gilbert equation: an overview},
  author={Lakshmanan, Muthusamy},
  journal={Philosophical Transactions of the Royal Society A: Mathematical, Physical and Engineering Sciences},
  volume={369},
  number={1939},
  pages={1280--1300},
  year={2011},
  publisher={The Royal Society Publishing}
}

@article{song2021direct,
  title={Direct visualization of magnetic domains and moir{\'e} magnetism in twisted 2D magnets},
  author={Song, Tiancheng and Sun, Qi-Chao and Anderson, Eric and Wang, Chong and Qian, Jimin and Taniguchi, Takashi and Watanabe, Kenji and McGuire, Michael A and St{\"o}hr, Rainer and Xiao, Di and others},
  journal={Science},
  volume={374},
  number={6571},
  pages={1140--1144},
  year={2021},
  publisher={American Association for the Advancement of Science}
}

@article{tan2021domain,
  title={Domain-wall confinement and dynamics in a quantum simulator},
  author={Tan, Wen Lin and Becker, Patrick and Liu, F and Pagano, G and Collins, KS and De, A and Feng, L and Kaplan, HB and Kyprianidis, A and Lundgren, R and others},
  journal={Nature Physics},
  volume={17},
  number={6},
  pages={742--747},
  year={2021},
  publisher={Nature Publishing Group UK London}
}

@article{matsubara2004computer,
  title={Computer simulation of magnetic domains in ultrathin magnetic films with a few monolayers},
  author={Matsubara, F and Endoh, S},
  journal={Journal of magnetism and magnetic materials},
  volume={272},
  pages={679--680},
  year={2004},
  publisher={Elsevier}
}

@article{kaappa2024magnetic,
  title={Magnetic domain walls interacting with dislocations in micromagnetic simulations},
  author={Kaappa, Sami and Santa-aho, Suvi and Honkanen, Mari and Vippola, Minnamari and Laurson, Lasse},
  journal={Communications Materials},
  volume={5},
  number={1},
  pages={256},
  year={2024},
  publisher={Nature Publishing Group UK London}
}

@article{de2023skyrmion,
  title={Skyrmion motion in magnetic anisotropy gradients: Acceleration caused by deformation},
  author={de Assis, Ismael Ribeiro and Mertig, Ingrid and G{\"o}bel, B{\"o}rge},
  journal={Physical Review B},
  volume={108},
  number={14},
  pages={144438},
  year={2023},
  publisher={APS}
}

@article{tazes2024efficient,
  title={Efficient Magnetic Vortex Acceleration by femtosecond laser interaction with long living optically shaped gas targets in the near critical density plasma regime},
  author={Tazes, I and Passalidis, S and Kaselouris, E and Mancelli, D and Karvounis, C and Skoulakis, A and Fitilis, I and Bakarezos, M and Papadogiannis, NA and Dimitriou, V and others},
  journal={Scientific Reports},
  volume={14},
  number={1},
  pages={4945},
  year={2024},
  publisher={Nature Publishing Group UK London}
}

@article{xie2025emerging,
  title={Emerging ferromagnetic materials for electrical spin injection: towards semiconductor spintronics},
  author={Xie, Yufang and Zhang, Su-Yun and Yin, Yin and Zheng, Naihang and Ali, Anwar and Younis, Muhammad and Ruan, Shuangchen and Zeng, Yu-Jia},
  journal={npj Spintronics},
  volume={3},
  number={1},
  pages={10},
  year={2025},
  publisher={Nature Publishing Group UK London}
}

@article{ahn20202d,
  title={2D materials for spintronic devices},
  author={Ahn, Ethan C},
  journal={npj 2D Materials and Applications},
  volume={4},
  number={1},
  pages={17},
  year={2020},
  publisher={Nature Publishing Group UK London}
}

@article{sousa2005non,
  title={Non-volatile magnetic random access memories (MRAM)},
  author={Sousa, Ricardo C and Prejbeanu, I Lucian},
  journal={Comptes Rendus Physique},
  volume={6},
  number={9},
  pages={1013--1021},
  year={2005},
  publisher={Elsevier}
}

@article{yuan2021extremely,
  title={Extremely high sensitivity magnetic field sensing based on birefringence-induced dispersion turning point characteristics of microfiber coupler},
  author={Yuan, Min and Pu, Shengli and Li, Dihui and Li, Yongxi and Hao, Zijian and Zhang, Yuxiu and Zhang, Chencheng and Yan, Shaokang},
  journal={Results in Physics},
  volume={29},
  pages={104743},
  year={2021},
  publisher={Elsevier}
}

@article{AMReX_JOSS,
  doi = {10.21105/joss.01370},
  url = {https://doi.org/10.21105/joss.01370},
  year = {2019},
  month = may,
  publisher = {The Open Journal},
  volume = {4},
  number = {37},
  pages = {1370},
  author = {Weiqun Zhang and Ann Almgren and Vince Beckner and John Bell and Johannes Blaschke and Cy Chan and Marcus Day and Brian Friesen and Kevin Gott and Daniel Graves and Max Katz and Andrew Myers and Tan Nguyen and Andrew Nonaka and Michele Rosso and Samuel Williams and Michael Zingale},
  title = {{AMReX}: a framework for block-structured adaptive mesh refinement},
  journal = {Journal of Open Source Software}
}

@article{gardner2022sundials,
  title     = {Enabling new flexibility in the {SUNDIALS} suite of nonlinear and differential/algebraic equation solvers},
  author    = {Gardner, David J and Reynolds, Daniel R and Woodward, Carol S and Balos, Cody J},
  journal   = {ACM Transactions on Mathematical Software (TOMS)},
  publisher = {ACM},
  volume    = {48},
  number    = {3},
  pages     = {1--24},
  year      = {2022},
  doi       = {10.1145/3539801}
}

@article{hindmarsh2005sundials,
  title     = {{SUNDIALS}: Suite of nonlinear and differential/algebraic equation solvers},
  author    = {Hindmarsh, Alan C and Brown, Peter N and Grant, Keith E and Lee, Steven L and Serban, Radu and Shumaker, Dan E and Woodward, Carol S},
  journal   = {ACM Transactions on Mathematical Software (TOMS)},
  publisher = {ACM},
  volume    = {31},
  number    = {3},
  pages     = {363--396},
  year      = {2005},
  doi       = {10.1145/1089014.1089020}
}

@article{li2020fourier,
  title={Fourier neural operator for parametric partial differential equations},
  author={Li, Zongyi and Kovachki, Nikola and Azizzadenesheli, Kamyar and Liu, Burigede and Bhattacharya, Kaushik and Stuart, Andrew and Anandkumar, Anima},
  journal={arXiv preprint arXiv:2010.08895},
  year={2020}
}

@article{sandu2019class,
  title = {A class of multirate infinitesimal GARK methods},
  author = {Sandu, Adrian},
  year = {2019},
  month = jan,
  volume = {57},
  pages = {2300--2327},
  issn = {0036-1429, 1095-7170},
  doi = {10.1137/18M1205492},
  journal = {SIAM Journal on Numerical Analysis},
  number = {5}
}

@article{chinomona2021implicit,
	title = {Implicit-explicit multirate infinitesimal {GARK} methods},
	volume = {43},
	url = {https://epubs.siam.org/doi/10.1137/20M1354349},
	doi = {10.1137/20M1354349},
	language = {en},
	number = {5},
	urldate = {2021-10-06},
	journal = {SIAM Journal on Scientific Computing},
	author = {Chinomona, Rujeko and Reynolds, Daniel R.},
	month = jan,
	year = {2021},
	pages = {A3082--A3113},
}

@article{yao2025advancing,
  title         = {Advancing simulations of coupled electron and phonon nonequilibrium dynamics using adaptive and multirate time integration},
  author        = {Jia Yao and Ivan Maliyov and David J. Gardner and Carol S. Woodward and Marco Bernardi},
  journal       = {npj Computational Materials},
  volume        = {11},
  number        = {1},
  pages         = {256},
  year          = {2025},
  doi           = {10.1038/s41524-025-01738-8}
}


%









\end{document}